\documentclass[sn-vancouver,Numbered]{sn-jnl}% Math and Physical Sciences Reference Style

\usepackage{graphicx}%
\usepackage{multirow}%
\usepackage{amsmath,amssymb,amsfonts}%
\usepackage{amsthm}%
\usepackage{mathrsfs}%
\usepackage{float}
\usepackage[title]{appendix}%
\graphicspath{{./Figures/}}
\usepackage{subcaption}
\usepackage{xcolor}%
\usepackage{tabularx}
\usepackage{tabularray}
\usepackage{qcircuit}
\usepackage{textcomp}%
\usepackage{manyfoot}%
\usepackage{braket}
\usepackage{colortbl}
\usepackage{dcolumn}
\usepackage{dblfloatfix}
\usepackage[section]{placeins}
\usepackage{booktabs}%
\usepackage{algorithm}
\usepackage{algorithmicx}%
\usepackage{algpseudocode}
\usepackage{listings}%
\begin{document}

\title[Article Title]{Quantum-Inspired Optimization Process for Data Imputation}

\author*[1]{\fnm{Nishikanta} \sur{Mohanty}}\email{nishikanta.m.mohanty@alumni.uts.edu.au}

\author[2,3]{\fnm{Bikash K.} \sur{Behera}}\email{bikas.riki@gmail.com}

\author[4]{\fnm{Badshah} \sur{Mukherjee}}\email{badshah.mukherjee@outlook.com}

\author[5]{\fnm{Pravat} \sur{Dash}}\email{pravat.dash@outlook.com}

\author[1]{\fnm{Christopher} \sur{Ferrie}}\email{Christopher.Ferrie@uts.edu.au}

\affil[1]{\orgdiv{Centre for Quantum Software and Information}, \orgname{University of Technology Sydney}, \orgaddress{\street{15 Broadway, Ultimo}, \city{Sydney}, \postcode{2007}, \state{NSW}, \country{Australia}}}

\affil[2]{\orgdiv{} \orgname{Bikash's Quantum (OPC) Pvt. Ltd.}, \orgaddress{\street{Balindi}, \city{Mohanpur}, \postcode{741246}, \state{WB}, \country{India}}}

\affil[3]{\orgdiv{} \orgname{Universit\`a degli Studi di Cagliari}, \orgaddress{\city{Cagliari}, \postcode{09123}, \country{Italy}}}

\affil[4]{\orgdiv{} \orgname{Independent Researcher}, \orgaddress{ \city{Dubai}, \postcode{9262}, \country{UAE}}}

\affil[5]{\orgdiv{} \orgname{Independent Researcher}, \orgaddress{ \city{Bangalore}, \postcode{560068}, \country{India}}}

\abstract{Data imputation is a critical preprocessing step, particularly in clinical datasets where missingness can bias downstream inference. We propose a quantum-inspired imputation framework and evaluate it on the \emph{UCI Pima Indians Diabetes} dataset, where several physiological variables contain structural zeros that are treated as missing. The approach combines Principal Component Analysis (PCA) with quantum rotation-based reconstruction, where rotation angles are selected via gradient-free optimizers (Simulated Annealing, COBYLA, and Differential Evolution). For robustness, we retain the top three optimizer solutions per method and reconstruct missing values under distribution-aware bounds (e.g., $\pm 2$ standard deviations) and avoid unrealistic collapse toward central tendency. We also benchmark against (i) classical baselines (Mean, KNN, MICE, Decision Tree), (ii) modern matrix/ML imputers (SoftImpute, MissForest, VAE), and (iii) rotation/PCA controls (random orthogonal rotations and bootstrap PCA). Across the evaluated features, quantum-inspired variants achieve substantially improved distributional fidelity under Wasserstein W1 relative to modern baselines (e.g., the best quantum setting reduces average W1 by $\sim$66\% vs.\ SoftImpute, $\sim$84\% vs.\ MissForest, and $\sim$90\% vs.\ VAE; and by $\sim$60\% vs.\ Mean imputation), while remaining competitive with strong classical/control configurations. Dispersion is better preserved under the quantum-inspired methods, with variance/SD ratios closer to the observed data than (SoftImpute, MissForest, and VAE) and typically better than (Mean, KNN, MICE, and Decision Tree) on average. Overall, the results support quantum-inspired, optimization-guided rotations as a statistically grounded and scalable strategy for realistic imputation in healthcare and general AI pipelines, with a design that can be mapped to future quantum hardware as it matures.}


\maketitle

\section{Introduction}
Data imputation is a statistical technique for addressing missing or partial data values within a dataset. Missing data may arise from various sources, including sensor faults, human errors, system failures, or privacy constraints \cite{LittleRubin2020}. The imputation process replaces missing values with estimates derived from the available data while preserving the dataset's integrity and minimizing bias \cite{Rubin1987}.
Imputation plays a vital role in numerous sectors and scenarios where data completeness is essential for analysis and decision-making. For instance, in healthcare and medicine, missing values in patient health records are prevalent due to inconsistent data entry or unavailable measurements. Imputation methods can fill gaps in laboratory test results, vital signs, or prescription histories, thereby supporting diagnostic and predictive modeling \cite{SchaferGraham2002}. Similarly, in finance and banking, datasets often have missing transaction details or customer profiles caused by system failures or incomplete submissions. Addressing these gaps through imputation enhances the reliability of credit scoring, risk analysis, and fraud detection algorithms \cite{Bertsimas2018}.
By employing appropriate imputation approaches tailored to the specific dataset and domain, organizations can ensure data completeness, improve model performance, and extract meaningful insights despite the presence of missing values. Advanced imputation methods, such as machine learning-based techniques, have further demonstrated their efficacy in handling complex datasets and mitigating the impact of missing data \cite{VanBuuren2018}. Given the increasing reliance on high-quality datasets in healthcare, finance, and AI pipelines, improving imputation accuracy with robust statistical adherence is not just a technical challenge but a foundational requirement for downstream decision-making.

\subsection{Quantum Imputation}
Imputation remains a critical focus of research in machine learning, with numerous classical algorithms developed to address missing data. These techniques range from simple statistical methods, such as mean, median, and mode imputation, to more sophisticated approaches like k-Nearest Neighbors (KNN) imputation, regression-based methods, and Bayesian imputation. Each of these methods is tailored to specific scenarios of missingness, including Missing Completely at Random (MCAR), Missing at Random (MAR), and Missing Not at Random (MNAR). Recent advancements suggest that quantum processes hold the potential to introduce innovative techniques for data imputation. Notable studies have begun exploring the use of quantum circuits for this purpose \cite{Kazdaghli_2023, Kochan_Zhang_Yang_2022, Sanavio_IEEE_2024}. Building on these developments, this paper proposes a novel quantum-classical hybrid method for data imputation that primarily addresses MAR scenarios while offering adaptability to MNAR cases through domain-specific adjustments. The proposed framework combines quantum-assisted rotations with classical principal component analysis (PCA) to achieve robust and statistically sound imputation. However, conventional techniques often fail to introduce realistic variability in the imputed values, leading to artificially smooth or biased reconstructions. To address this, we propose a quantum-inspired imputation framework that achieves statistically diverse and more representative imputations.

The method begins by imputing missing values in the dataset using a measure of central tendency, such as mean, median, or mode. PCA is then applied to reduce the dimensionality of the dataset, capturing the essential variance within a smaller set of principal components. These components are subjected to quantum-assisted rotations, optimized through a systematic process, to generate rotated versions of the principal components. The rotated components are subsequently reconstructed into the original feature space, where their averaged values replace the initial central tendency imputations. A key feature of this method is the enforcement of statistical bounds, ensuring that reconstructed values remain within $\pm 2$ standard deviations (SD) of the original data distribution. In addition, a penalty mechanism is integrated to discourage values from clustering excessively around the imputed central tendency measure, thereby preserving the natural variability in the data. This quantum-classical imputation framework represents a significant advancement in data preprocessing, leveraging the unique properties of quantum rotations to address missing data challenges more effectively. Our approach improves distributional fidelity at the dataset level across evaluated features, the best quantum-inspired configurations achieve mean Wasserstein W1 on the order of $\sim 10$-$13$ (with weaker baselines reaching $\sim 17$+), while simultaneously preserving dispersion more realistically. In particular, the average deviation in dispersion ratios drops to $\approx 0.38$ for variance ($|Var\_{imp}/Var\_{obs}-1|$) and $\approx 0.22$ for standard deviation ($|SD\_{imp}/SD\_{obs}-1|$), compared to $\approx 0.54/0.43$ for classical baselines and $\approx 1.08/0.59$ for modern matrix/ML imputers performed on the \emph{UCI Pima Indians Diabetes} dataset.

\subsection{Novelty and Contribution}
This study presents a novel quantum-assisted approach to data imputation, integrating quantum rotations with classical Principal Component Analysis (PCA) to enhance statistical integrity and computational efficiency. Unlike conventional methods, which often rely solely on classical optimization techniques, the proposed framework introduces quantum-inspired transformations to improve missing data reconstruction. The key contributions of this work are as follows:

\begin{itemize}
    \item \textbf{Quantum-Assisted Imputation:} We propose a data imputation framework that leverages quantum-inspired optimization, specifically utilizing quantum rotations and low-depth circuit representations to enhance the imputation process.
    
    \item \textbf{Hybrid Quantum-Classical Methodology:} Unlike previous quantum-assisted imputation techniques \cite{Kazdaghli_2023, Kochan_Zhang_Yang_2022, Sanavio_IEEE_2024}, which operate predominantly in quantum domains, this work integrates quantum transformations with classical PCA, ensuring adaptability for practical machine learning applications.
    
    \item \textbf{Statistical Constraint Enforcement:} The imputed values are constrained within statistically valid boundaries, ensuring adherence to the original dataset's mean and $\pm 2$ standard deviations, thereby preserving data consistency and avoiding extreme deviations.
    
    \item \textbf{Optimization Across Multiple Classical Algorithms:} To enhance computational efficiency and robustness, the framework evaluates multiple classical optimization techniques, including COBYLA, Differential Evolution, and Simulated Annealing, identifying the most effective approach for optimizing quantum-assisted imputation.
    
    \item \textbf{Scalability and Practical Implications:} The methodology is designed to be computationally scalable, making it suitable for real-world applications in high-dimensional datasets, including healthcare analytics, financial modeling, and IoT-based sensor data imputation.

    \item \textbf{Different from Modern ML Imputers (SoftImpute, MissForest, VAE) and classical analogues (Random Ortho, Bootstrap PCA):}
Unlike modern imputers that rely on a \emph{model class assumption} (e.g., low-rank matrix completion in SoftImpute, tree-ensemble predictors in MissForest, or neural generative reconstruction in VAE), our method is \emph{model-free} and instead searches for an \emph{optimization-guided rotation} in a PCA-compressed latent space. The rotation parameters form a low-dimensional, physically interpretable transformation (aligning with a low-depth circuit-style rotation view), and are selected by gradient-free optimizers under explicit distribution-aware constraints (bounded within statistically valid ranges), rather than learned via predictive training. Further, random orthogonal rotations and bootstrap PCA serve as \emph{controls} that apply transformations without an objective-driven angle selection; in contrast, our rotations are \emph{chosen to minimize distributional mismatch and dispersion distortion}, isolating the gain from optimization-guided structure rather than generic linear mixing.

\end{itemize}

\section{Process Flow and algorithms}\label{Algorithms}
Handling missing data is a fundamental challenge in data-driven applications. Traditional imputation methods are based on statistical measures that often introduce bias and fail to capture the complex interdependencies among features \cite{Rubin_1976_missingdata}. In this paper, we propose a quantum-enhanced imputation pipeline leveraging PCA and quantum rotation techniques. Our methodology ensures the imputed values remain consistent with the data distribution, minimizing information loss. The proposed approach is divided into six key steps, detailed below.

\subsection{Algorithm 1: Imputation of Missing Values}
In the beginning, we start with a data set with missing values and to address missing values in the dataset, we employ a selected central tendency measure - mean, median or mode - depending on the nature of the data \cite{Li_Stuart_Allison_2015}. This imputation ensures numerical completeness and enables PCA to be applied over the entire dataset \cite{Jolliffe_PCA_2002}. The consistency in imputation supports the stability of PCA transformation while preserving the statistical characteristics of the data. This approach maintains the interpretability of principal components and ensures that imputed values do not significantly distort the distribution.

\begin{algorithm}[H]
\caption{Impute Missing Values}
\label{Initial_imputation}
\begin{algorithmic}[1]
\State \textbf{Input:} Dataset $D$, columns to impute $C$, imputation method $M$
\For{each column $c \in C$}
    \If{$M$ is `mean`}
        \State Compute $v_c \leftarrow$ mean of nonzero values in $c$
    \ElsIf{$M$ is `median`}
        \State Compute $v_c \leftarrow$ median of nonzero values in $c$
    \ElsIf{$M$ is `mode`}
        \State Compute $v_c \leftarrow$ mode of nonzero values in $c$
    \Else
        \State Throw error ``Invalid method"
    \EndIf
    \State Replace 0s in $c$ with $v_c$
\EndFor
\State \textbf{Output:} Imputed dataset $D'$
\end{algorithmic}
\end{algorithm}

\subsection{Algorithm 2: Calculation of Statistics for Imputed Records}
Once missing values are imputed, their influence on the dataset is quantified. This step (Algorithm \ref{Calculate_Stats}) computes the central tendency and standard deviation of each imputed feature, creating a statistical reference point \cite{Donders_gentle_intro_missing_2006}. These computed statistics are later utilized in the inverse transformation step to ensure the corrections do not introduce statistical drift.

\begin{algorithm}[H]
\caption{Calculate Statistics for Imputed Records}
\label{Calculate_Stats}
\begin{algorithmic}[1]
\State \textbf{Input:} Dataset $D$, original dataset $D_o$, imputed columns $C$
\State Initialize statistics list $S$
\For{each row $r \in D$}
    \For{each column $c \in C$}
        \If{$D_o[r, c] = 0$}
            \State Compute central tendency $T_c \gets$ median of nonzero values in $c$
            \State Compute standard deviation $\sigma_c$
            \State Append $(r, c, T_c, T_c + 2\sigma_c, T_c - 2\sigma_c)$ to $S$
        \EndIf
    \EndFor
\EndFor
\State \textbf{Output:} Statistics dataset $S$
\end{algorithmic}
\end{algorithm}

\subsection{Algorithm 3, 4: PCA Selection and Optimized Quantum PCA Rotation}
PCA is a well-established method for dimensionality reduction \cite{Wold_Esbensen_Geladi_PCA_1987}; however, we enhance its effectiveness by leveraging quantum state representations. In our approach, PCA-transformed feature vectors are encoded as quantum states, enabling parameterised quantum rotations. Each feature vector undergoes a unitary transformation facilitated by controlled phase rotation gates, followed by classical state extraction. The rotation angle is dynamically optimized to ensure that transformed values remain aligned with the underlying data distribution, thus preserving structural integrity while enhancing feature representation. Algorithms \ref{choose pca components} and \ref{Rotate_PCA_Components} are created for the purpose while the former ensured we chose the number of PCA components as the power of 2 to aid in circuit formulation, and the later rotates the PCA feature vectors using a controlled phase rotation $R_x(\theta)$ on an encoded PCA feature vector.

\begin{algorithm}[H]
\caption{Choose PCA Components}
\label{choose pca components}
\begin{algorithmic}[1]
\State \textbf{Input:} Number of features $n$
\State Compute nearest power of 2: $p \leftarrow 2^{\lfloor \log_2(n) \rfloor}$
\State \textbf{Output:} $\max(2, p)$
\end{algorithmic}
\end{algorithm}

\begin{algorithm}[H]
\caption{Hybrid Quantum Rotation for PCA Components}
\label{Rotate_PCA_Components}
\begin{algorithmic}[1]
\State \textbf{Input:} Normalized PCA dataset $Z'$, rotation angle $\theta$, number of components $n$
\For{each row $Z'_i \in Z'$}
    \State Encode $Z'_i$ as a quantum state $\ket{\psi_i}$
    \State Initialize quantum circuit with $n$ qubits
    \For{each qubit $q \in \{1, \dots, n\}$}
        \State Apply controlled phase rotation $R_x(\theta)$ to qubit $q$
    \EndFor
    \State Apply quantum measurement and extract classical rotated values
    \State Store rotated values in $Z^*$
\EndFor
\State Normalize $Z^*$ and apply post-processing corrections
\State \textbf{Output:} Optimized rotated dataset $Z^*$
\end{algorithmic}
\end{algorithm}

\subsection{Algorithm 5: Optimized Inverse PCA Transformation and Correction}
After performing  parameterised quantum rotation of PCA components, an inverse transformation is required to map the transformed features back to their original space. However, conventional inverse PCA methods often produce unrealistic values \cite{Hastie_staistical_learning_2009}, particularly when imputed elements are disproportionately influenced by the initially applied central tendency measure. To mitigate this issue, we introduce a correction mechanism that leverages precomputed statistical properties. Specifically, features exhibiting significant deviations from the original distribution either exceeding the Interquartile Range (IQR) or falling beyond $\pm2SD$ are subjected to adaptive penalty adjustments. Additionally, a secondary penalty mechanism is applied to prevent imputed values from regressing toward the statistical references computed in Algorithm~\ref{Calculate_Stats}, effectively eliminating initial imputation bias. This dual adjustment strategy ensures that the reconstructed dataset remains statistically consistent with the original distribution while minimizing distortions introduced by the imputation process. The complete procedure for inverse transformation and penalty adjustments is outlined in Algorithm~\ref{Inverse_PCA}.

\medskip
\noindent\textbf{Quantum-inspired parameterised rotation (operator form).}
Let $z\in\mathbb{R}^{d}$ denote a PCA-latent row. After padding/normalisation, $z$ is amplitude-encoded into an $n$-qubit state $\ket{\psi(z)}\in\mathbb{C}^{2^n}$. Notably, although we apply a product-unitary rotation layer, the \emph{encoded state itself} can exhibit non-separability: in general $\ket{\psi(z)}$ need not admit a factorisation $\ket{\psi(z)}=\bigotimes_{q=1}^{n}\ket{\psi_q}$, i.e., amplitude encoding may implicitly represent entanglement through correlations across amplitudes. We then apply a low-depth layer of parameterised single-qubit rotations:
\begin{equation}
U(\Theta) \;=\; \bigotimes_{q=1}^{n} U_q(\theta_q),
\qquad
U_q(\theta_q)\in\{R_x(\theta_q),\,R_y(\theta_q),\,R_z(\theta_q)\},
\label{eq:product_unitary_rotations_alg5}
\end{equation}
yielding the rotated latent representation $Z^*$ that is subsequently mapped back to the original feature space via the inverse PCA step and corrected using the penalty rules below. We intentionally avoid introducing explicit entangling gates within the rotation layer (e.g., CNOT/CZ-based patterns) as a design choice: this maintains a compact, low-depth parameterisation and reduces sensitivity to spurious couplings and noise amplification that can arise when entanglement is actively generated by the rotation circuit, particularly under near-term execution or noisy simulation settings.

\begin{figure}[H]
\centering
\[
\Qcircuit @C=1.0em @R=0.9em {
\lstick{q_0} & \multigate{4}{\textsf{AmpEnc}(\lvert\psi(z)\rangle)} & \gate{R_x(\theta_0)} & \qw \\
\lstick{q_1} & \ghost{\textsf{AmpEnc}(\lvert\psi(z)\rangle)}       & \gate{R_x(\theta_1)} & \qw \\
\lstick{q_2} & \ghost{\textsf{AmpEnc}(\lvert\psi(z)\rangle)}       & \gate{R_x(\theta_2)} & \qw \\
\lstick{q_3} & \ghost{\textsf{AmpEnc}(\lvert\psi(z)\rangle)}       & \gate{R_x(\theta_3)} & \qw \\
\lstick{q_4} & \ghost{\textsf{AmpEnc}(\lvert\psi(z)\rangle)}       & \gate{R_x(\theta_4)} & \qw
}
\]
\caption{Illustrative low-depth circuit used to parameterise PCA-latent perturbations via amplitude encoding followed by single-qubit rotations.}
\label{fig:pca_rotation_circuit_alg5}
\end{figure}

\begin{algorithm}[H]
\caption{Hybrid Inverse PCA and Post-Processing Adjustment}
\label{Inverse_PCA}
\begin{algorithmic}[1]
\State \textbf{Input:} Rotated dataset $Z^*$, PCA model $P$, statistics dataset $S$
\State Denormalize $Z^*$ by multiplying with normalization factors
\State Compute inverse PCA transformation: $X' \gets Z^* U^T$
\For{each feature $f \in F$ and record $x'_i \in X'$}
    \State Retrieve original mean $\mu_f$ and standard deviation $\sigma_f$ from $S$
    \If{$x'_i$ falls outside $[\mu_f - 2\sigma_f,\ \mu_f + 2\sigma_f]$}
        \State Apply corrective shift using adaptive penalty function
    \EndIf
\EndFor
\State Apply final statistical adjustments based on prior distributions
\State \textbf{Output:} Fully corrected dataset $X^*$
\end{algorithmic}
\end{algorithm}

\subsection{Algorithm 6,7: Optimized Gradient-Free Rotation Angle Optimization}
Determining the best rotation angle(s) for quantum PCA transformation is key to achieving accurate feature reconstruction. Since traditional gradient-based approaches struggle with the complexities of high-dimensional quantum spaces \cite{Biamonte_QML_2017}, we take an alternative approach by employing gradient-free optimization techniques. Specifically, our method selects either \textbf{Differential Evolution} \cite{Storn_Price_DE_1997}, \textbf{Constrained Optimization BY Linear Approximations (COBYLA)} \cite{Powell_COBYLA_1994}, and \textbf{Simulated Annealing} \cite{Kirkpatrick_simulated_annealing_1988}. These algorithms efficiently search for optimal rotation angles without requiring explicit gradient calculations, making them well-suited for quantum-encoded PCA transformations. The optimization process is structured into two key components: \textbf{Algo: \ref{Angle_Cost_Function}} and \textbf{Algo: \ref{Optimization_Process}}. The first component defines a cost function that quantifies deviation based on the selected rotation angle, while the second applies a gradient-free optimization strategy to minimize this deviation over the normalized PCA components. Instead of selecting a single absolute optimal angle, we instruct the optimization algorithm to retain the three best angles with the lowest deviation. This approach provides robustness against potential variations in the dataset while ensuring stability in the imputation process.

\begin{algorithm}[H]
\caption{Compute Cost Function for Optimization}
\label{Angle_Cost_Function}
\begin{algorithmic}[1]
\State \textbf{Input:} Rotation angle $\theta$, PCA dataset $Z$, statistics dataset $S$
\State Rotate PCA dataset using $\theta$
\State Compute inverse PCA transform and apply penalty adjustments
\State Compute total deviation from the original dataset
\State \textbf{Output:} Cost value
\end{algorithmic}
\end{algorithm}

\begin{algorithm}[H]
\caption{Hybrid Gradient-Free Optimization for Quantum Rotation}
\label{Optimization_Process}
\begin{algorithmic}[1]
\State \textbf{Input:} Cost function $f(\theta)$, PCA dataset $Z$, statistics dataset $S$, bounds $[0, \pi]$
\State Initialize optimizer with a diverse population of angles
\For{each iteration until convergence}
    \State Evaluate cost function $f(\theta)$ using dataset $Z$ and statistics $S$
    \State Apply gradient-free strategy (e.g., differential evolution, COBYLA, simulated annealing) to explore solution space
    \State Update candidate solutions based on cost minimization
    \State Select top-performing angles for the next iteration
\EndFor
\State \textbf{Output:} Optimized rotation angles $\theta^*$
\end{algorithmic}
\end{algorithm}

\subsection{Algorithm 8,9: Quantum Imputation Pipeline}
The final step integrates all preceding methods into a cohesive pipeline, ensuring a structured and efficient approach to quantum-enhanced imputation. The process unfolds as follows:

\begin{enumerate}
    \item Missing values are imputed using a selected central tendency measure, ensuring data completeness.
    \item The statistical properties of imputed features, such as central tendency and variance, are computed and stored as reference points for later correction.
    \item PCA is applied, and the data undergoes a quantum rotation to improve feature representation and distribution alignment.
    \item The inverse PCA transformation is carried out, incorporating penalty adjustments to maintain statistical integrity and prevent unrealistic reconstructions.
    \item Using a gradient-free optimization approach, the optimal rotation angles are identified and refined.
    \item Instead of selecting a single optimal angle, the algorithm saves the three best angles, each producing a separate rotated dataset.
    \item The final dataset is generated by averaging the values from the three rotated datasets for each imputed feature. These averaged values are then superimposed onto the original dataset in place of the imputed values. The statistical reference points from earlier steps serve as validation criteria to ensure the correctness of the final imputed dataset.
\end{enumerate}

This pipeline offers a novel approach to handling missing data by integrating statistical imputation, quantum-enhanced transformations, and gradient-free optimization, providing a robust and adaptive solution for improving dataset integrity. The algorithms \ref{Average_Results} and \ref{Quantum_Imputation_pipeline} detail the steps for the final imputation process.

\begin{algorithm}[H]
\caption{Average Optimized Results}
\label{Average_Results}
\begin{algorithmic}[1]
\State \textbf{Input:} Transformed datasets $Z_1$, $Z_2$, $Z_3$
\State Compute average dataset $Z' \gets (Z_1 + Z_2 + Z_3) / 3$
\State \textbf{Output:} Averaged dataset $Z'$
\end{algorithmic}
\end{algorithm}

\begin{algorithm}[H]
\caption{Quantum Imputation Pipeline}
\label{Quantum_Imputation_pipeline}
\begin{algorithmic}[1]
\State \textbf{Input:} Dataset $D$, PCA model, imputation columns $C$
\State Perform the following steps:
\State \hspace{1em} Step 1: Impute missing values
\State \hspace{1em} Step 2: Calculate statistics for imputed values
\State \hspace{1em} Step 3: Select PCA components dynamically
\State \hspace{1em} Step 4: Apply optimized quantum PCA rotation
\State \hspace{1em} Step 5: Perform inverse PCA transformation and penalty adjustment
\State \hspace{1em} Step 6: Compute the angle cost function
\State \hspace{1em} Step 7: Optimize rotation angle using gradient-free methods
\State \hspace{1em} Step 8: Average multiple optimized outputs
\State \hspace{1em} Step 9: Save the final corrected dataset
\State \textbf{Output:} Corrected dataset $D'$
\end{algorithmic}
\end{algorithm}

\begin{figure*}
     \centering
    \includegraphics[width=\linewidth]{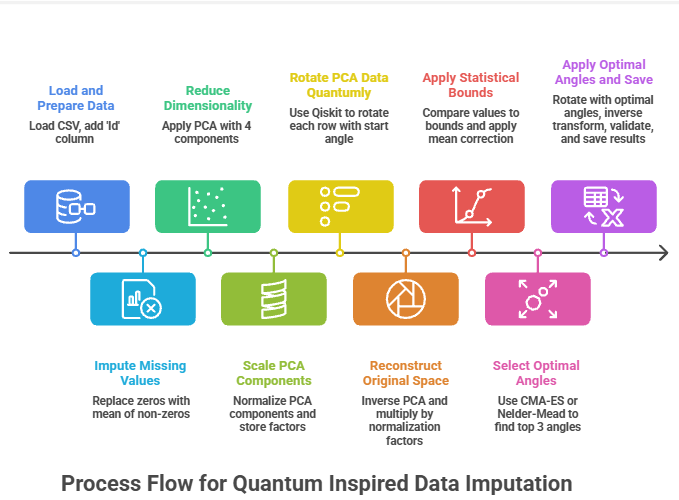}
    \caption{Overall Imputation Process Flow}
    \label{Imputation_process_flow}
\end{figure*}

\section{Experimental Setup, data gathering, and statistics}
We started our experiment by choosing a public dataset ``Diabetes patient records" from the UCI Machine learning repository \cite{Daibetis_UCI_ML_Repo}. This is a simple dataset with 9 columns, with one column being an output column. The dataset was purposefully chosen due to its high percentage of missing values. 

\begin{figure}[H]
    \centering
    \includegraphics[width=0.7\linewidth]{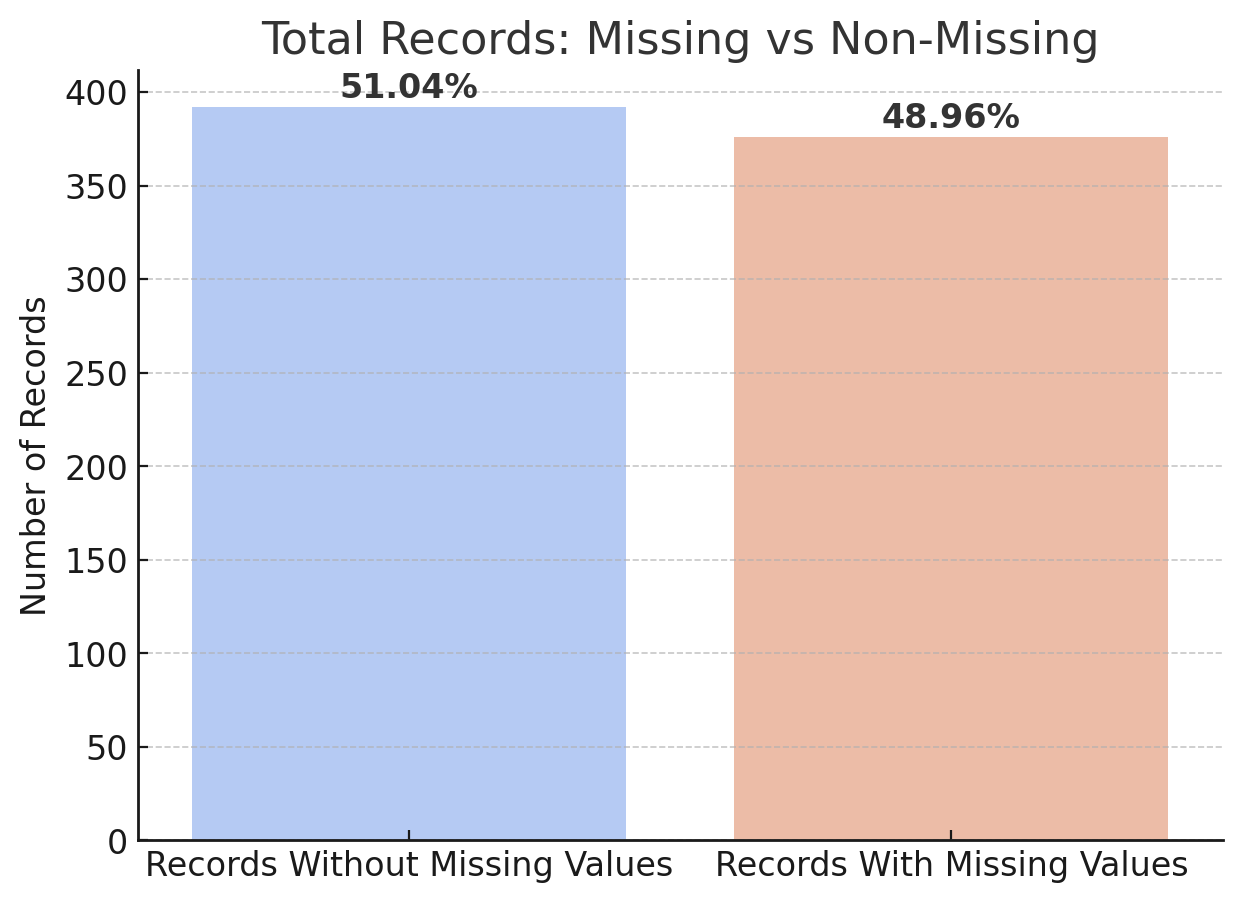}
    \caption{Missing vs Nonmissing values recordwise}
    \label{Missing values}
\end{figure}

\begin{figure}[H]
    \centering
    \includegraphics[width=0.7\linewidth]{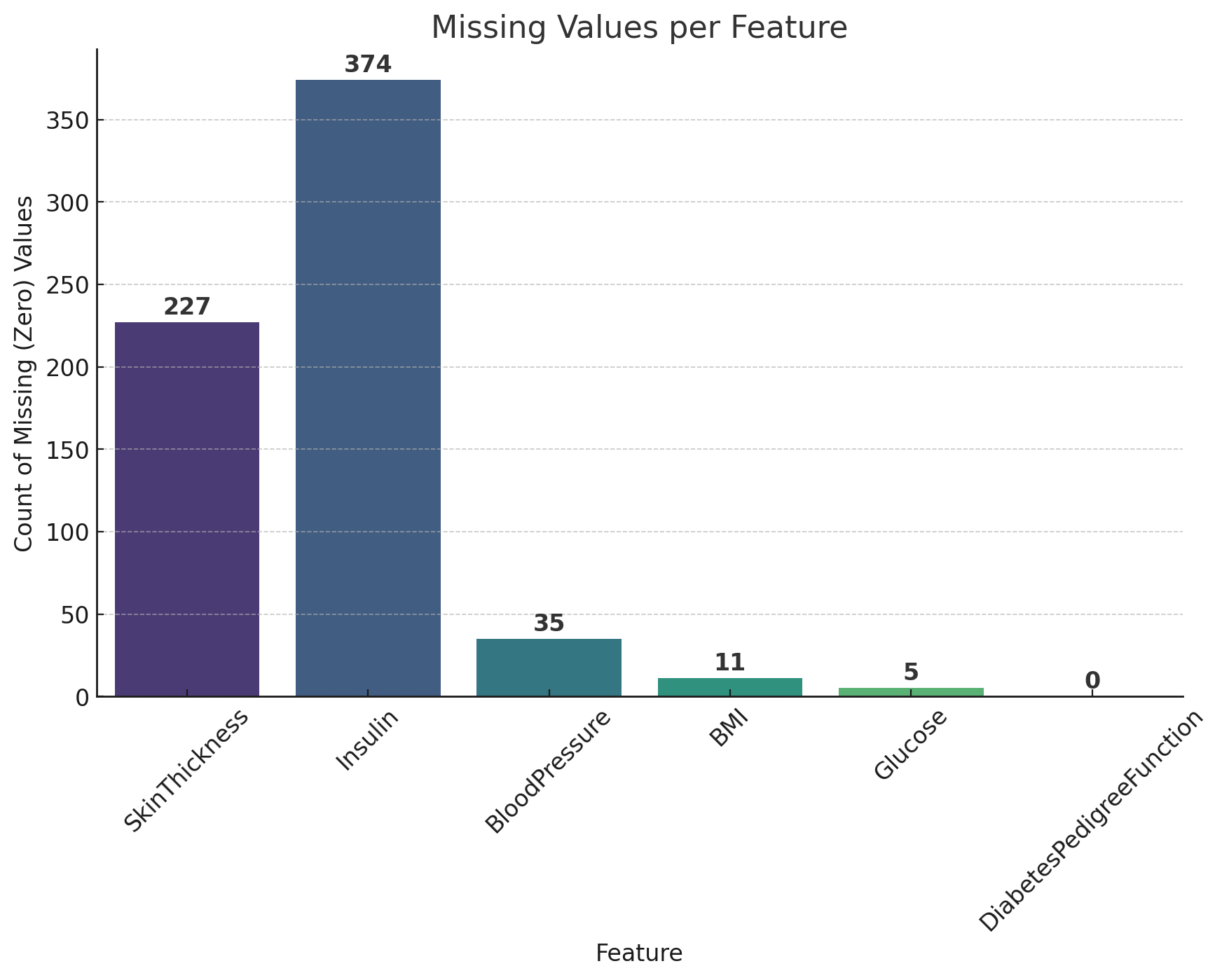}
    \caption{Featurewise Missing values }
    \label{Feature_Missing}
\end{figure}

Out of the nine features available in the diabetes dataset \cite{Daibetis_UCI_ML_Repo}, we identified six variables containing missing or zero-valued entries: `SkinThickness', `Insulin', `BloodPressure', `BMI', `Glucose', and `DiabetesPedigreeFunction'. These were chosen for imputation as they exhibited biologically improbable zero values, which likely indicate missing data. Conversely, the features `Outcome' and `Pregnancies' were excluded from the imputation process. `Outcome' serves as the target variable, making it unsuitable for modification. Similarly, `Pregnancies' was retained as-is since a zero value in this feature represents a valid, natural condition (i.e., no prior pregnancies). Therefore, only the selected six features were subjected to quantum imputation techniques to reconstruct their missing values while maintaining data integrity. The dataset lacked an ID column, which we have created as part of the process; this column helps identify records and aids in optimization as well as the rotation process. To assess the effectiveness of quantum-based imputation, we processed the dataset through our quantum imputation pipeline and compared its performance against three conventional imputation techniques: Mean, KNN, MICE, and Decision Tree Imputation. The quantum imputation approach incorporated three distinct gradient-free optimization algorithms: Differential Evolution \cite{Storn_Price_DE_1997}, COBYLA \cite{Powell_COBYLA_1994}, and Simulated Annealing \cite{Kirkpatrick_simulated_annealing_1988}. This comparative analysis allowed us to examine the efficiency and accuracy of quantum-driven optimization in reconstructing missing values while maintaining statistical coherence with the original dataset. We will detail these evaluation metrics in the following section.

\section{Results} \label{Results}
A rigorous statistical evaluation is required to assess the quality of data imputation, particularly for the proposed quantum-inspired imputation framework. In this study, we compare classical baselines (Mean, MICE, KNN, Decision Tree, ) against quantum-inspired optimization methods (Simulated Annealing, COBYLA, and Differential Evolution). Our primary evaluation criteria are distributional similarity and dispersion realism, quantified using the Kolmogorov--Smirnov (KS) test, Wasserstein (W1) distance, variance, and standard deviation (SD).As described in the previous section, the quantum-inspired pipeline produces multiple optimized rotated datasets by retaining the three best rotation angles for each optimizer (Annealing, COBYLA, and Differential Evolution). These rotated outputs are treated as the quantum-imputed candidates. We then benchmark them against classical imputations by (i) comparing the empirical distributions of \emph{imputed-only} values to the corresponding \emph{observed} values for each feature, and (ii) computing KS, W1, variance, and SD statistics to quantify how closely each method preserves the underlying data characteristics. The following subsections present the distributional comparisons and summarize the statistical evidence derived from these metrics. In subsequent headings, we will detail distributions and draw inferences from statistical measures.

\subsection{Distribution of Features}
After applying the rotation process, we analyzed the distribution of the imputed features, comparing them against the original dataset and across different quantum-based optimization methods. This analysis provides insight into how the reconstruction mechanisms and penalty functions influence the final rotated dataset post-optimization.

\subsubsection{Glucose}

\begin{figure}[H]
    \centering
    \includegraphics[width=0.7\linewidth]{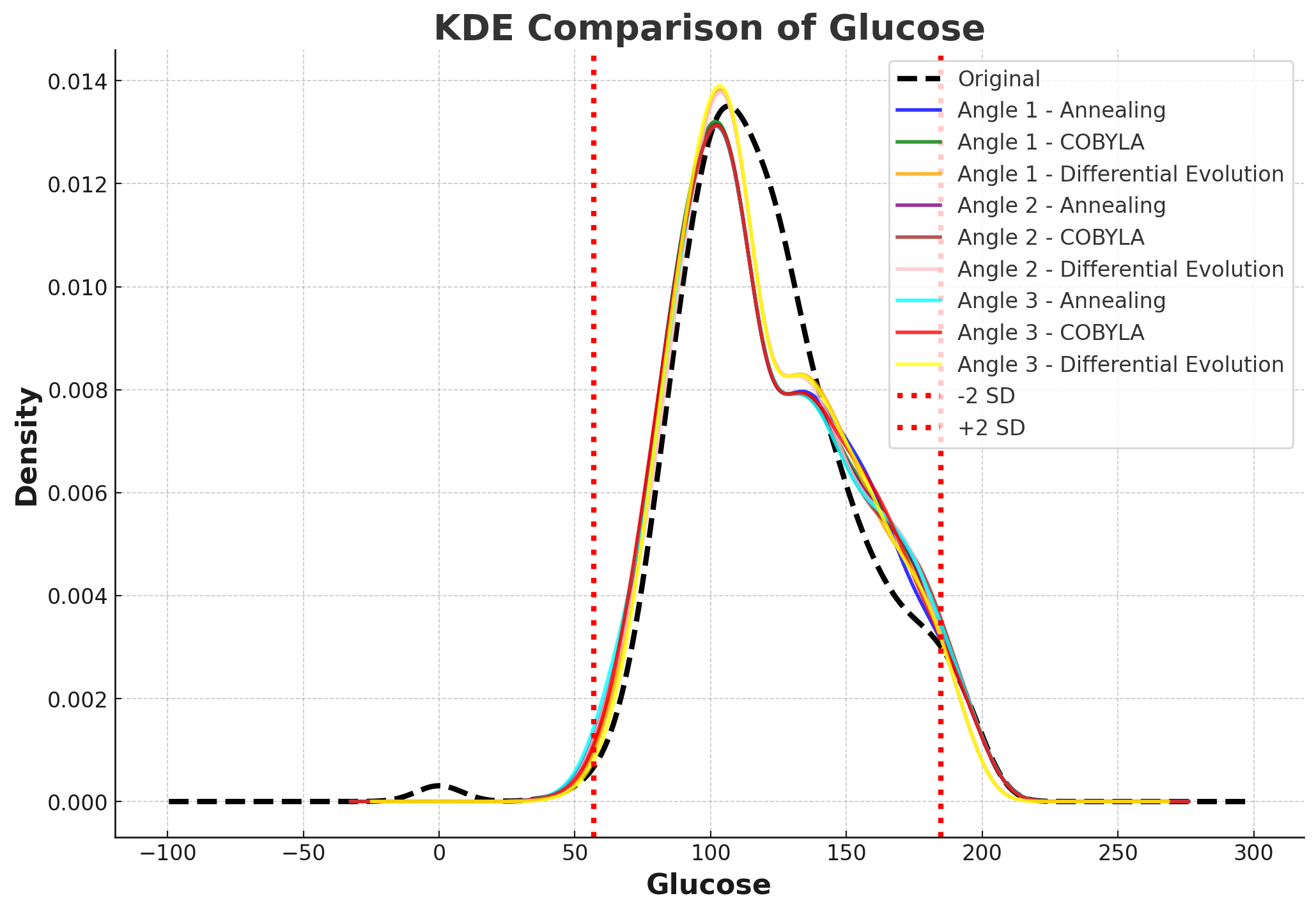}
    \caption{KDE of Glucose: Original vs. Rotated Datasets.}
    \label{fig:Glucose_distribution}
\end{figure}

The KDE plots of Glucose \ref{fig:Glucose_distribution} for the rotated datasets exhibit a strong overlap with the original distribution in the primary range of approximately 60 to 180. A key distinction is the absence of 0 values in the rotated data, while the original dataset included instances of glucose levels at 0, all rotated datasets start around 44-49, effectively removing the artificial spike at 0. As a result, the lower tail of the distribution is truncated above 0, aligning with the lower bound of approximately 55 (close to the original mean – $2\sigma$). The density curves exhibit a single peak between 100-130, remaining well within the original $\pm2\sigma$ interval. The mean value (~121) is marked by a dashed line, with the $\pm2\sigma$ bounds (~57 and ~185) indicated by dotted lines in the plot. \textbf{(Plot title: KDE of Glucose: Original vs. Rotated Datasets. \ref{fig:Glucose_distribution})}

\subsubsection{Blood Pressure}

\begin{figure}[H] 
    \centering
    \includegraphics[width=0.7\linewidth]{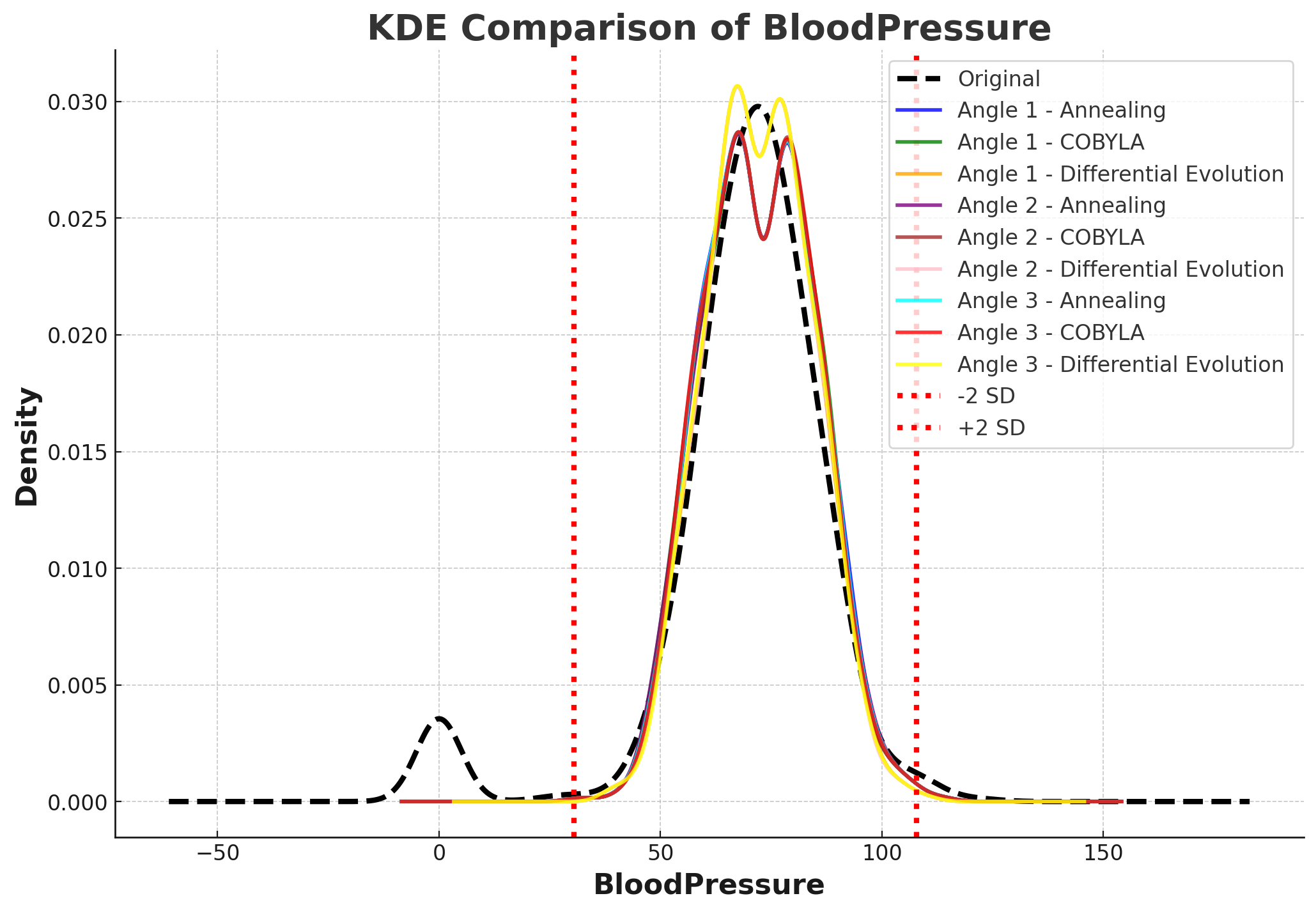}
    \caption{KDE of BloodPressure: Original vs. Rotated Datasets}
    \label{fig:BloodPressure_distribution}
\end{figure}

KDE plots of Blood Pressure \ref{fig:BloodPressure_distribution}, the rotated datasets exhibit distributions very similar to the original for normal blood pressure values. Notably, all rotated datasets lack 0 values; the minimum blood pressure in the rotated datasets is approximately 32, eliminating the original dataset's sharp peak at 0. Instead, the distribution begins near 30, aligning with the lower bound of the original distribution (~30.4, just above $-2\sigma$). The main body of the rotated distributions (approximately 50-90 mmHg) closely follows the original dataset's density curve. Additionally, the upper bound of the rotated curves is generally around 110, aligning with the original dataset's upper bound (~108, near $+2\sigma$). Reference lines at ~69 (mean) and ~30/108 ($\pm2\sigma$) provide statistical context for the visual representation. \textbf{(Plot title: KDE of BloodPressure: Original vs. Rotated Datasets \ref{fig:BloodPressure_distribution})}

\subsubsection{Skin Thickness}

\begin{figure}[H] 
    \centering
    \includegraphics[width=0.7\linewidth]{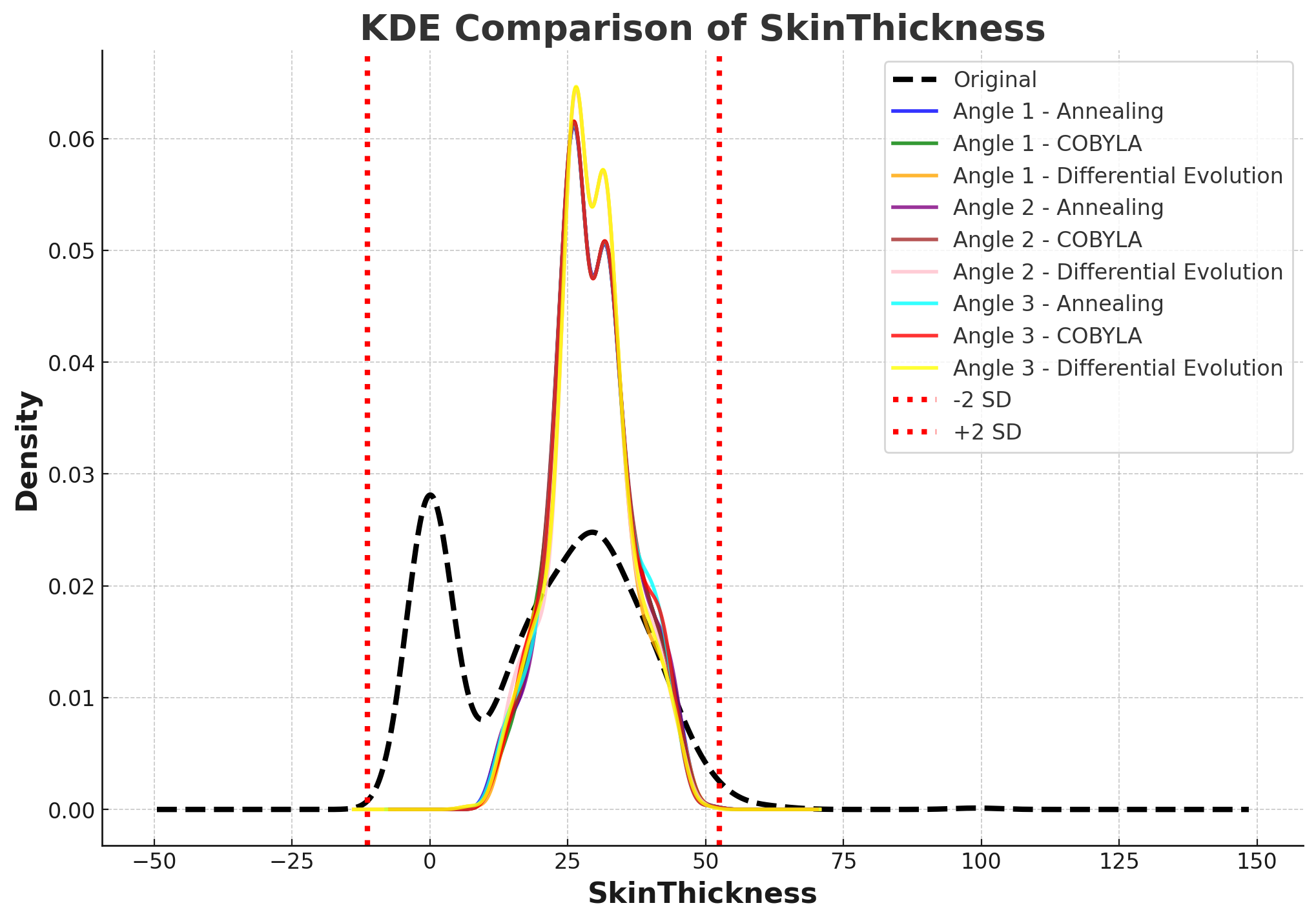}
    \caption{KDE of Skin Thickness: Original vs. Rotated Datasets}
    \label{fig:SkinThickness_distribution}
\end{figure}

The original dataset contains numerous 0 values for SkinThickness, indicative of missing data, resulting in a pronounced peak at 0 and a long tail extending towards higher values (up to ~99). The KDE plots \ref{fig:SkinThickness_distribution} for the rotated datasets reveal a much more constrained and realistic distribution. The rotation process effectively eliminates the artificial spike at 0, with minimum values ranging between 7 and 12 across different optimization methods. Instead of clustering at 0, the rotated distributions exhibit a peak between 20 and 30, closely resembling the distribution of observed (non-missing) values in the original dataset. Additionally, the rotated datasets truncate the right tail, capping maximum values around 50 (approximately the original mean $+2\sigma$ at ~52.4). While the original dataset contained extreme outliers approaching 100, the rotated distributions remain well within the $\pm2\sigma$ range, ensuring a more natural data spread. \textbf{(Plot title: KDE of Skin Thickness: Original vs. Rotated Datasets \ref{fig:SkinThickness_distribution})}

\subsubsection{Insulin}

\begin{figure}[H]
    \centering
    \includegraphics[width=0.7\linewidth]{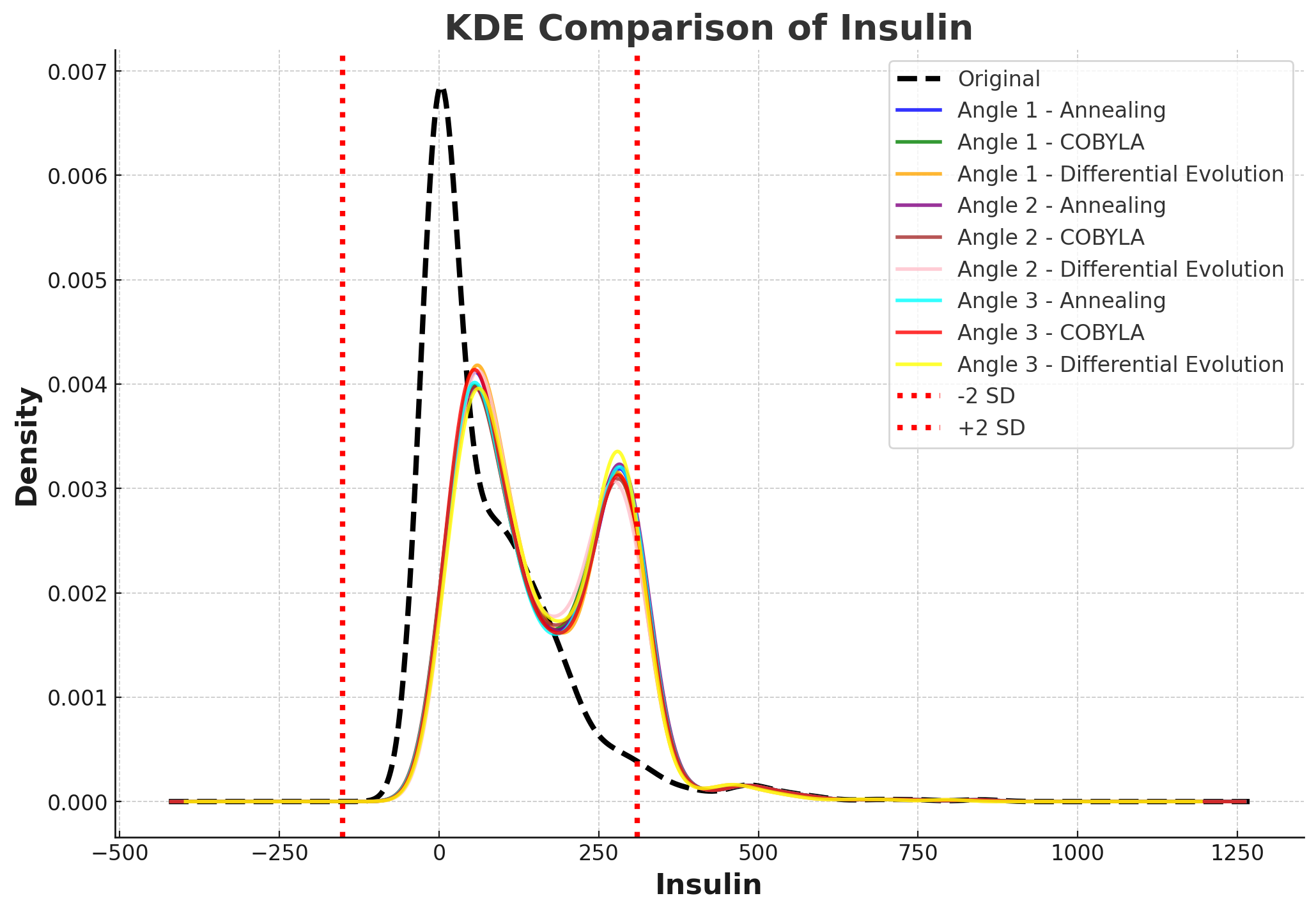}
    \caption{KDE of Insulin: Original vs. Rotated Datasets}
    \label{fig:Insulin_distribution}
\end{figure}

The original insulin distribution is highly skewed due to a substantial proportion of 0 values (representing missing data) and a long right tail extending up to 846. The KDE \ref{fig:Insulin_distribution} plots of the rotated datasets show a similar right-skewed distribution but without the artificial peak at 0. Most rotated datasets replaced missing values with small positive values (ranging from ~0.2 to 1.2), substantially reducing the density spike at 0. Consequently, the rotated distributions exhibit a smoother, more realistic distribution, with a peak in the tens to low hundreds. On the higher end, the rotated datasets tend to cap extreme values at approximately 795–841, slightly below the original dataset's maximum (846), and with much lower density at extreme values. Nearly the entire density of each rotated dataset remains within the $\pm2\sigma$ range (~0 to 310). \textbf{(Plot title: KDE of Insulin: Original vs. Rotated Datasets.\ref{fig:Insulin_distribution})}

\subsubsection{BMI}

\begin{figure} 
    \centering
    \includegraphics[width=0.7\linewidth]{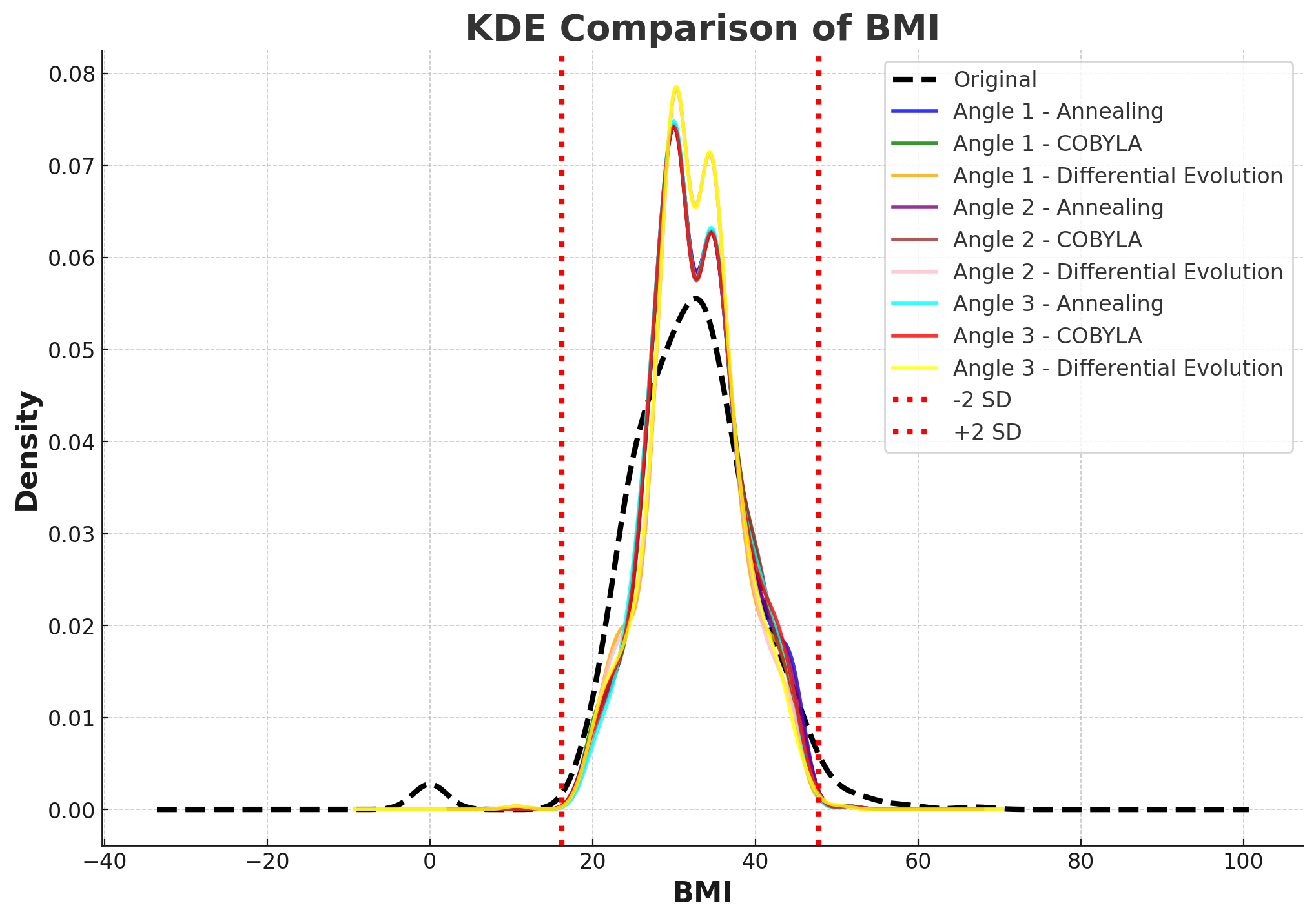}
    \caption{KDE of BMI: Original vs. Rotated Datasets}
    \label{fig:BMI_distribution}
\end{figure}

The original BMI distribution is approximately bell-shaped around 32 but contains a significant proportion of 0 values, introducing left-skew. Additionally, some extreme BMI values reach up to 67. The KDE \ref{fig:BMI_distribution} plots for the rotated datasets show a high degree of similarity to the original dataset but with key refinements at the extremes. The 0 values are removed, with minimum BMI values now starting between 10 and 18 across optimization methods. This adjustment eliminates the artificial mass at 0 and ensures a more realistic distribution. The overall shape remains similar, with peaks around 30 and values ranging between ~20 and 50, aligning closely with the original $\pm2\sigma$ range (~16.2 to 47.8). This suggests that the rotation process effectively prevents extreme values while preserving the dataset’s natural variance. \textbf{(Plot title: KDE of BMI: Original vs. Rotated Datasets.\ref{fig:BMI_distribution})}

\subsubsection{Diabetes Pedigree Function}

\begin{figure} 
    \centering
    \includegraphics[width=0.7\linewidth]{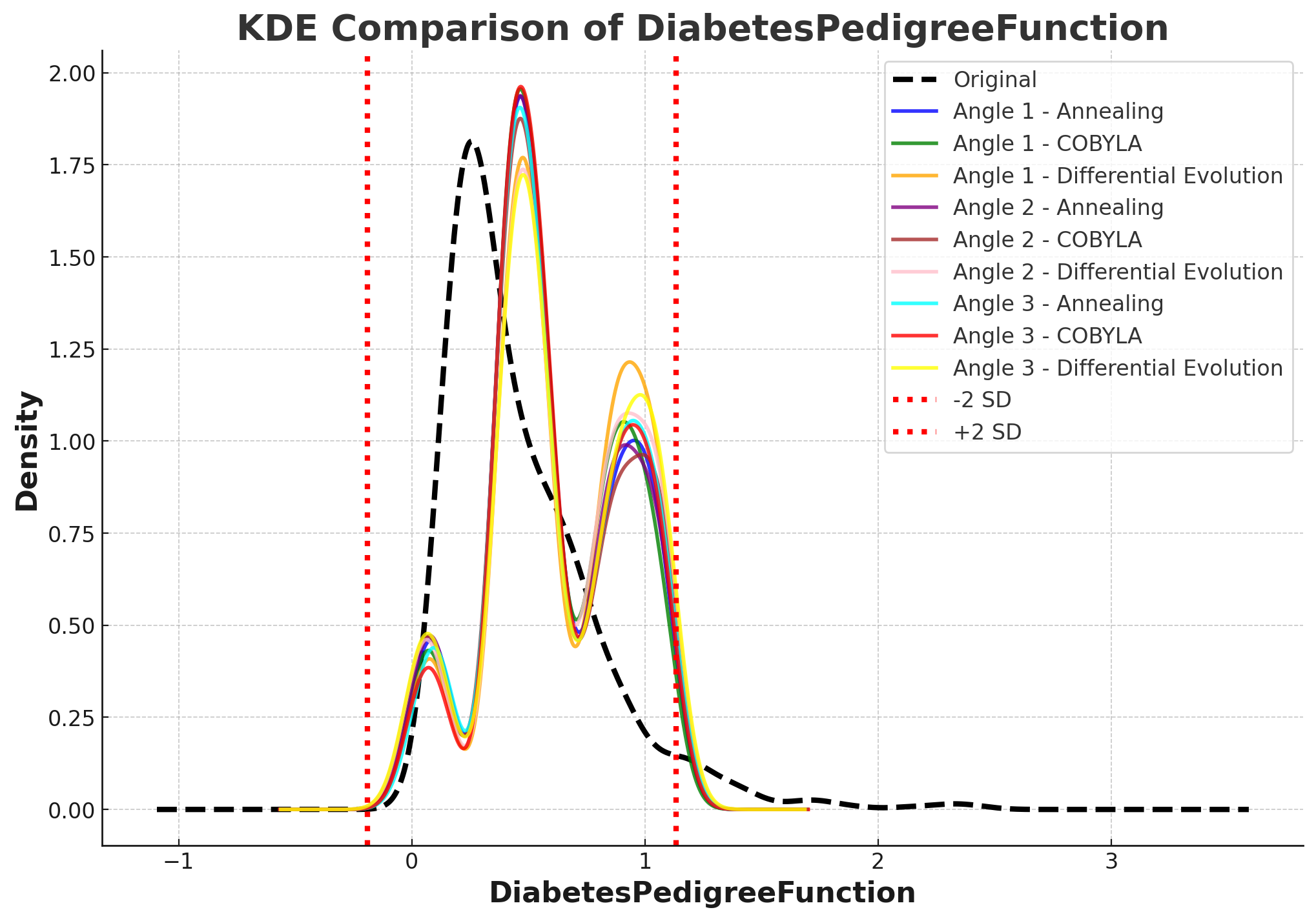}
    \caption{KDE of Diabetes Pedigree Function: Original vs. Rotated Datasets}
    \label{fig:DiabetesPedigree_distribution}
\end{figure}

The Diabetes Pedigree Function (DPF) is a useful baseline for comparison as it contains no missing values and was not subject to mean imputation. The original dataset exhibits a right-skewed distribution, with most values between 0 and 1 and some outliers extending up to 2.42. The KDE \ref{fig:DiabetesPedigree_distribution} plots reveal that the rotated datasets largely preserve the original DPF distribution, demonstrating that the transformation method does not unnecessarily alter features with naturally occurring variability. However, a key difference is that none of the rotated datasets contain values exceeding 1.13, effectively capping the distribution at the original $+2\sigma$ threshold. This ensures that extreme outliers are removed while maintaining the overall feature structure. \textbf{(Plot title: KDE of Diabetes Pedigree Function: Original vs. Rotated Datasets.\ref{fig:DiabetesPedigree_distribution})}

\subsubsection{Observations from Distribution Analysis}

\begin{itemize}
\item \textbf{Enhanced Variability in Mean-Imputed Features:} The optimizer-rotation process successfully diversified the imputed features (SkinThickness, Insulin, BloodPressure, BMI, Glucose) by redistributing values that were previously concentrated at the mean. This adjustment ensures that formerly identical missing-value imputed points now exhibit a spread that aligns with natural variability.
\item \textbf{Preserving Realistic Data Boundaries:} The rotated imputed values remain well within the $\pm2\sigma$ range for each feature, preventing extreme outliers while avoiding artificial constraints that would overly restrict feature variability.
\item \textbf{Minimal Impact on Non-Imputed Features:} Diabetes Pedigree Function, which was not subject to imputation, retained its natural distribution, confirming that the optimization primarily affects features impacted by missing data rather than indiscriminately modifying all variables.
\item \textbf{Comparison of Optimization Methods:} Simulated Annealing and COBYLA produced highly similar KDE distributions, suggesting convergence to a stable imputation model. Differential Evolution introduced slightly greater variance in glucose and insulin distributions but still adhered to the overall statistical constraints.
\item \textbf{Retention of Population Characteristics:} The final rotated dataset maintains the integrity of the original dataset while mitigating the biases introduced by mean imputation, ensuring that distributions remain faithful to real-world data patterns.
\end{itemize}

\subsection{Kolmogorov-Smirnov (KS) Test}

\begin{figure*}[!t]
    \centering
    \begin{subfigure}{\textwidth}
        \centering
        \includegraphics[width=\linewidth]{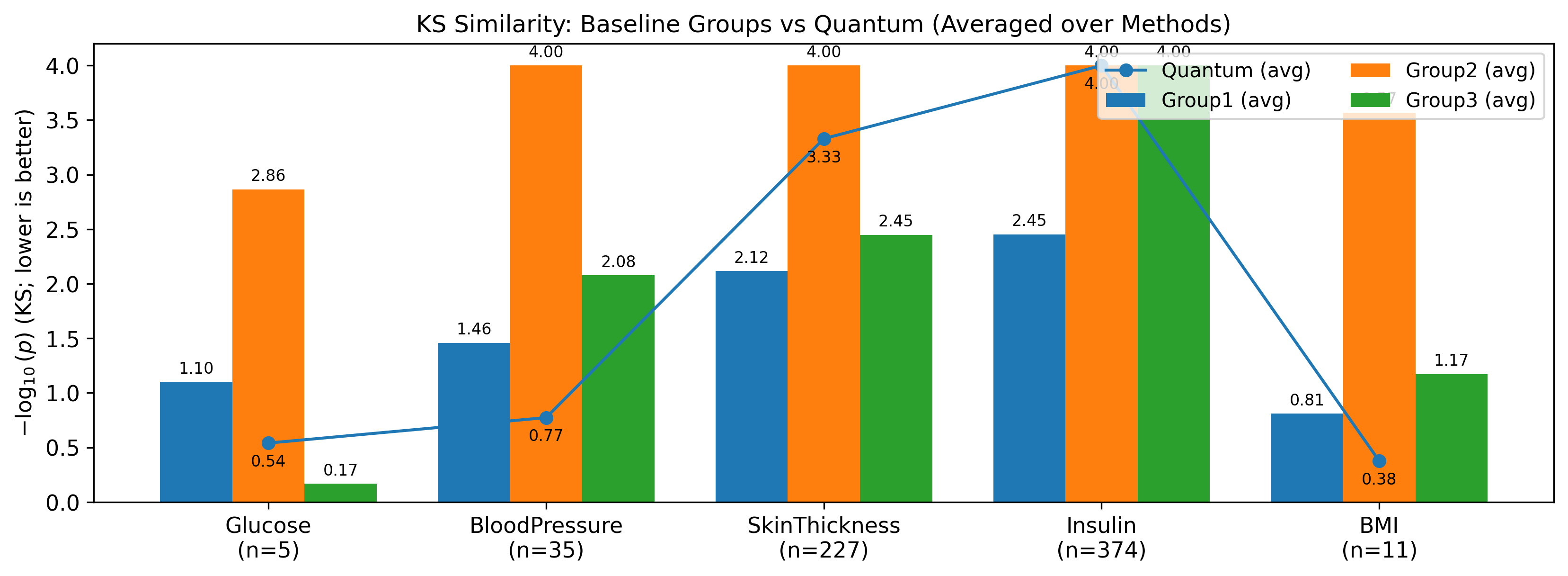}
        \caption{KS p-values across features (group averages as bars; quantum average shown as a line). Higher is better overlap.}
        \label{fig:KS_abs_barchart}
    \end{subfigure}\hfill
    \begin{subfigure}{\textwidth}
        \centering
        \includegraphics[width=\linewidth]{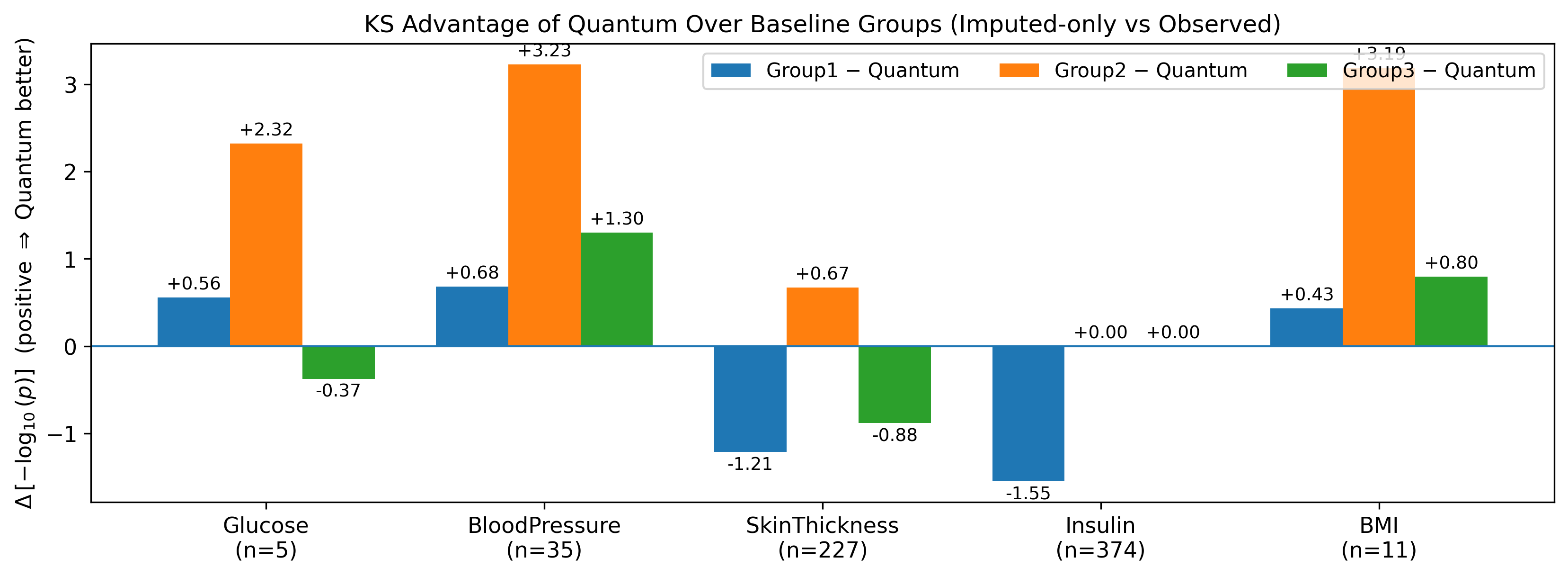}
        \caption{KS advantage of Quantum over baseline groups: $\Delta p = p_{\text{Quantum avg}} - p_{\text{Group avg}}$. Positive values favour Quantum.}
        \label{fig:KS_delta_barchart}
    \end{subfigure}
    \caption{Grouped KS test results (imputed-only vs observed).}
    \label{fig:KS_grouped}
\end{figure*}

The Kolmogorov--Smirnov (KS) test \cite{Massey_KSTest_1951} is a non-parametric test that compares two one-dimensional distributions by measuring the maximum absolute difference between their empirical cumulative distribution functions (ECDFs). In this study, we apply the KS test to compare the \emph{imputed-only} samples against the \emph{observed} samples for each feature, thereby assessing whether the imputation procedure induces a statistically detectable shift in the marginal distribution. A higher KS p-value indicates insufficient evidence to reject the null hypothesis that the two samples are drawn from the same distribution, while a lower p-value indicates a detectable distributional deviation. Because KS p-values can become highly sensitive under large missingness and moderate sample sizes, we interpret KS comparatively across method groups and corroborate conclusions with complementary metrics such as Wasserstein W1 (distributional distance) and dispersion realism (SD-ratio). The KS results are summarized in Table~\ref{tab:ks_all_groups_quantum} and visualized in Fig.~\ref{fig:KS_delta_barchart} and Fig.~\ref{fig:KS_abs_barchart}.

\FloatBarrier
\begin{table*}[!tb]
\centering
\caption{KS Test p-values (imputed-only vs.\ observed): Grouped baselines compared with Quantum imputation methods. $N_{\text{missing}}$ reports the number of missing (structural-zero) entries per feature.}
\label{tab:ks_all_groups_quantum}

\textbf{(a) Group 1 (Classical) vs Quantum}
\resizebox{1.2\textwidth}{!}{%
\begin{tabular}{l r rrrrrrrrrrrrr}
\toprule
Feature Name & $N_{\text{missing}}$ & KNN & MICE & Mean & DecisionTree & DE\_1 & DE\_2 & DE\_3 & COBYLA\_1 & COBYLA\_2 & COBYLA\_3 & Annealing\_1 & Annealing\_2 & Annealing\_3 \\
\midrule
SkinThickness & 227 & 0.002 & 0.024 & 0.000 & 0.714 & 0.000 & 0.003 & 0.000 & 0.000 & 0.001 & 0.009 & 0.004 & 0.000 & 0.000 \\
Insulin & 374 & 0.043 & 0.000 & 0.000 & 0.358 & 0.000 & 0.000 & 0.000 & 0.000 & 0.000 & 0.000 & 0.000 & 0.000 & 0.000 \\
BloodPressure & 35 & 0.054 & 0.329 & 0.000 & 0.835 & 0.408 & 0.444 & 0.217 & 0.353 & 0.032 & 0.491 & 0.012 & 0.080 & 0.520 \\
BMI & 11 & 0.250 & 0.917 & 0.003 & 0.837 & 0.608 & 0.754 & 0.223 & 0.608 & 0.207 & 0.114 & 0.554 & 0.830 & 0.616 \\
Glucose & 5 & 0.256 & 0.116 & 0.061 & 0.022 & 0.006 & 0.941 & 0.870 & 0.403 & 0.513 & 0.401 & 0.813 & 0.234 & 0.177 \\
\bottomrule
\end{tabular}%
}

\vspace{0.8em}
\textbf{(b) Group 2 (SoftImpute / MissForest / VAE) vs Quantum}
\resizebox{1.2\textwidth}{!}{%
\begin{tabular}{l r rrrrrrrrrrrr}
\toprule
Feature Name & $N_{\text{missing}}$ & SoftImpute & MissForest & VAE & DE\_1 & DE\_2 & DE\_3 & COBYLA\_1 & COBYLA\_2 & COBYLA\_3 & Annealing\_1 & Annealing\_2 & Annealing\_3 \\
\midrule
SkinThickness & 227 & 0.000 & 0.000 & 0.000 & 0.000 & 0.003 & 0.000 & 0.000 & 0.001 & 0.009 & 0.004 & 0.000 & 0.000 \\
Insulin & 374 & 0.000 & 0.000 & 0.000 & 0.000 & 0.000 & 0.000 & 0.000 & 0.000 & 0.000 & 0.000 & 0.000 & 0.000 \\
BloodPressure & 35 & 0.000 & 0.000 & 0.000 & 0.408 & 0.444 & 0.217 & 0.353 & 0.032 & 0.491 & 0.012 & 0.080 & 0.520 \\
BMI & 11 & 0.002 & 0.000 & 0.000 & 0.608 & 0.754 & 0.223 & 0.608 & 0.207 & 0.114 & 0.554 & 0.830 & 0.616 \\
Glucose & 5 & 0.000 & 0.257 & 0.000 & 0.006 & 0.941 & 0.870 & 0.403 & 0.513 & 0.401 & 0.813 & 0.234 & 0.177 \\
\bottomrule
\end{tabular}%
}

\vspace{0.8em}
\textbf{(c) Group 3 (Random Orthogonal Rotation / Bootstrap PCA) vs Quantum}
\resizebox{1.2\textwidth}{!}{%
\begin{tabular}{l r rrrrrrrrrrrr}
\toprule
Feature Name & $N_{\text{missing}}$ & RandOrtho\_Mean & RandOrtho\_MICE & BootPCA & DE\_1 & DE\_2 & DE\_3 & COBYLA\_1 & COBYLA\_2 & COBYLA\_3 & Annealing\_1 & Annealing\_2 & Annealing\_3 \\
\midrule
SkinThickness & 227 & 0.000 & 0.453 & 0.001 & 0.000 & 0.003 & 0.000 & 0.000 & 0.001 & 0.009 & 0.004 & 0.000 & 0.000 \\
Insulin & 374 & 0.000 & 0.000 & 0.000 & 0.000 & 0.000 & 0.000 & 0.000 & 0.000 & 0.000 & 0.000 & 0.000 & 0.000 \\
BloodPressure & 35 & 0.000 & 0.036 & 0.163 & 0.408 & 0.444 & 0.217 & 0.353 & 0.032 & 0.491 & 0.012 & 0.080 & 0.520 \\
BMI & 11 & 0.006 & 0.660 & 0.077 & 0.608 & 0.754 & 0.223 & 0.608 & 0.207 & 0.114 & 0.554 & 0.830 & 0.616 \\
Glucose & 5 & 0.610 & 0.892 & 0.579 & 0.006 & 0.941 & 0.870 & 0.403 & 0.513 & 0.401 & 0.813 & 0.234 & 0.177 \\
\bottomrule
\end{tabular}%
}
\end{table*}

\FloatBarrier

\subsubsection{KS test Observations (Grouped Baselines vs. Quantum):}
\begin{itemize}
    \item \textbf{Group 1 (Mean/KNN/MICE/DT) vs. Quantum:} Classical baselines are mixed under KS: while \textit{DecisionTree} can yield high p-values on some variables (e.g., \textit{BloodPressure} $\approx 0.835$; \textit{SkinThickness} $\approx 0.714$), simpler strategies (notably \textit{Mean}) frequently collapse to near-zero p-values on high-missingness features. In contrast, quantum variants remain \emph{consistently competitive} on low-to-moderate missingness variables, achieving strong KS alignment for \textit{Glucose} (e.g., \textit{DE\_2} $\approx 0.941$) and \textit{BMI} (e.g., \textit{Annealing\_2} $\approx 0.830$), indicating robust distribution preservation beyond central-tendency heuristics.

    \item \textbf{Group 2 (SoftImpute/MissForest/VAE) vs. Quantum:} Group~2 exhibits systematic KS failure, with p-values reported as \textbf{0.000} (rounded) across several key features (\textit{BloodPressure}, \textit{SkinThickness}, \textit{Insulin}), implying a strong detectable shift between \emph{imputed-only} and \emph{observed} distributions. Quantum optimization substantially mitigates this mismatch, delivering materially higher p-values across multiple variables (e.g., \textit{BloodPressure} up to $\approx 0.520$; \textit{Glucose} up to $\approx 0.941$), demonstrating clear superiority over modern matrix/ML baselines in preserving marginal distributions.

    \item \textbf{Group 3 (Random Orthogonal Rotation / Bootstrap PCA) vs. Quantum:} Rotation/PCA controls can match distributions for select low-missingness features (e.g., \textit{RandOrtho\_MICE} on \textit{Glucose} $\approx 0.892$), but they degrade sharply on more challenging variables (e.g., \textit{Insulin} $\rightarrow 0.000$; \textit{BloodPressure} best $\approx 0.163$). Quantum methods show broader robustness, including strong KS performance on \textit{BloodPressure} (e.g., \textit{Annealing\_3} $\approx 0.520$) while maintaining high p-values on \textit{Glucose}/\textit{BMI}, indicating that the gains are not explained by generic rotations alone.

    \item \textbf{Overall (KS): Quantum  as a group-level result.} Aggregating across features and optimizer variants, quantum methods provide the most reliable \emph{distribution-preserving} behavior relative to Group~2 and demonstrate competitive or superior alignment relative to Group~1/3 on the variables where KS is most informative (low-to-moderate missingness). For high-missingness features (\textit{Insulin}, \textit{SkinThickness}), KS becomes extremely sensitive; nevertheless, quantum remains a strong contender when interpreted jointly with W1 and dispersion realism, supporting quantum superiority at the \emph{group-level} rather than as a universal per-feature guarantee.
\end{itemize}

\subsection{Wasserstein Distance}
\begin{figure*}[!t]
    \centering
    \begin{subfigure}{\textwidth}
        \centering
        \includegraphics[width=\linewidth]{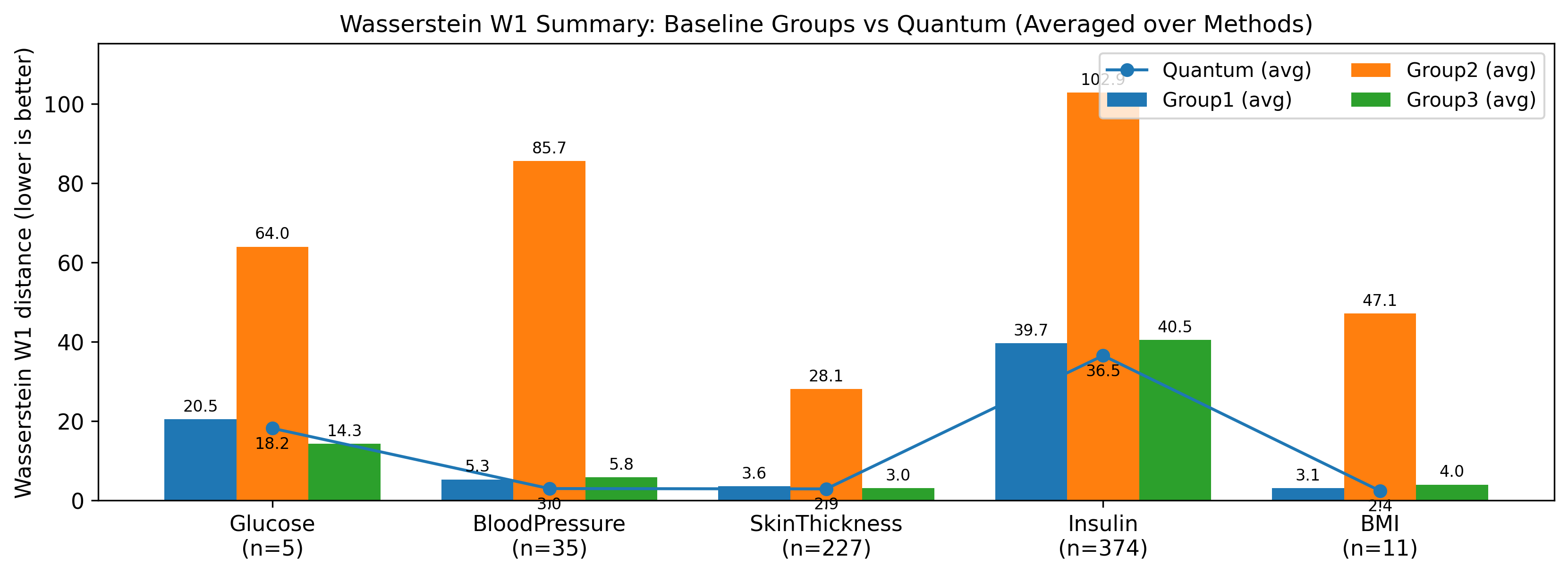}
        \caption{Wasserstein W1 distance across features (Group averages as bars; Quantum average shown as a line). Lower W1 indicates better distributional fidelity.}
        \label{fig:Wasserstein_W1_barchart}
    \end{subfigure}\hfill
    \begin{subfigure}{\textwidth}
        \centering
        \includegraphics[width=\linewidth]{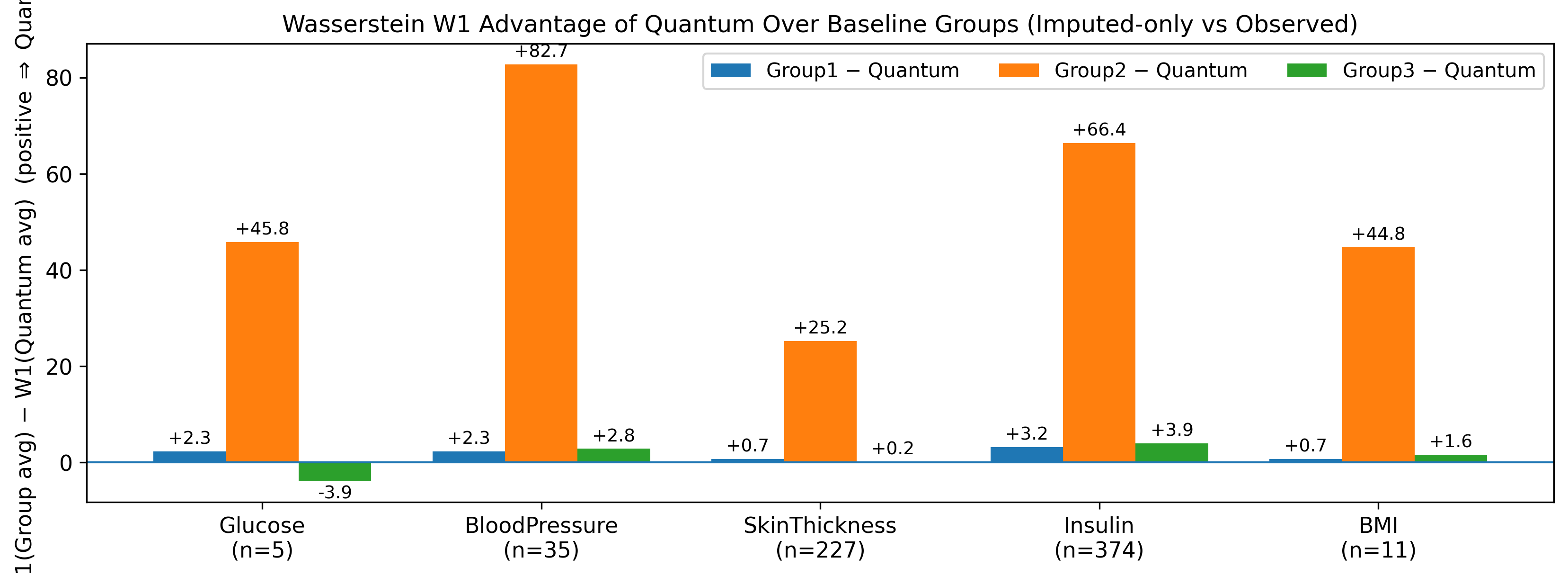}
        \caption{Wasserstein W1 advantage of Quantum over baseline groups across features. Bars report $\Delta$W1 $=$ W1(Group avg) $-$ W1(Quantum avg); positive values indicate Quantum achieves lower W1.}
        \label{fig:Wasserstein_W1_delta_barchart}
    \end{subfigure}
    \caption{Wasserstein W1 results (imputed-only vs.\ observed): absolute group comparisons and Quantum advantage.}
    \label{fig:Wasserstein_W1_combined}
\end{figure*}

Based on the idea of optimal transport, the Wasserstein distance \cite{Villani_Wasserstein_2008,Ramdas_Wasserstein_2017} measures the distance between two probability distributions by quantifying the minimum “work” required to transform one distribution into the other. In the context of imputation evaluation, Wasserstein W1 provides a direct, distance-based assessment of how closely the \emph{imputed-only} values match the \emph{observed} values for each feature. A smaller W1 indicates stronger distributional fidelity, i.e., the imputation procedure reconstructs missing-value distributions that remain closer to the empirical structure of the observed data.
In our experiments, we report W1 comparisons in a grouped manner: Group~1 (Mean/KNN/MICE/DecisionTree) vs quantum , Group~2 (SoftImpute/MissForest/VAE) vs quantum , and Group~3 controls (random orthogonal rotations and bootstrap PCA) vs quantum (Table~\ref{tab:w1_all_groups_quantum}). The corresponding visual summaries are shown in Figs.\ref{fig:Wasserstein_W1_barchart} and \ref{fig:Wasserstein_W1_delta_barchart}. Overall, the grouped view clarifies that quantum methods consistently reduce distributional transport distance relative to weaker baselines (especially Group~2) while remaining competitive with the strongest classical/control methods. W1 is also feature-dependent: high-missingness variables such as \textit{Insulin} and \textit{SkinThickness} typically exhibit larger transport distances across methods, reflecting the intrinsic difficulty of reconstructing their missing-value distributions. These results motivate the grouped observations discussed below.

\FloatBarrier

\begin{table*}[!tb]
\centering
\caption{Wasserstein W1 distances (imputed-only vs.\ observed): Grouped baselines compared with Quantum imputation methods. $N_{\text{missing}}$ reports the number of missing (structural-zero) entries per feature. Lower W1 indicates better distributional fidelity.}
\label{tab:w1_all_groups_quantum}

\textbf{(a) Group 1 (Classical) vs Quantum}\\
\resizebox{1.2\textwidth}{!}{%
\begin{tabular}{l r rrrrrrrrrrrrr}
\toprule
Feature Name & $N_{\text{missing}}$ & KNN & MICE & Mean & DecisionTree & DE\_1 & DE\_2 & DE\_3 & COBYLA\_1 & COBYLA\_2 & COBYLA\_3 & Annealing\_1 & Annealing\_2 & Annealing\_3 \\
\midrule
Glucose & 5 & 18.825 & 18.079 & 24.633 & 20.350 & 38.693 & 10.097 & 9.071 & 16.653 & 18.693 & 16.574 & 12.201 & 21.069 & 20.451 \\
BloodPressure & 35 & 4.408 & 4.050 & 9.614 & 3.023 & 2.546 & 2.444 & 3.193 & 2.927 & 4.405 & 2.970 & 3.637 & 2.410 & 2.069 \\
SkinThickness & 227 & 3.042 & 1.744 & 8.271 & 1.176 & 3.014 & 2.917 & 2.734 & 3.017 & 3.191 & 2.525 & 2.745 & 3.204 & 2.524 \\
Insulin & 374 & 24.561 & 37.073 & 82.880 & 14.208 & 38.957 & 35.906 & 34.760 & 38.044 & 35.260 & 34.376 & 37.404 & 36.081 & 37.976 \\
BMI & 11 & 3.040 & 1.380 & 5.462 & 2.398 & 2.095 & 2.023 & 3.154 & 2.190 & 3.028 & 2.612 & 2.220 & 2.145 & 1.956 \\
\bottomrule
\end{tabular}%
}

\vspace{0.8em}
\textbf{(b) Group 2 (SoftImpute / MissForest / VAE) vs Quantum}\\
\resizebox{1.2\textwidth}{!}{%
\begin{tabular}{l r rrrrrrrrrrrrr}
\toprule
Feature Name & $N_{\text{missing}}$ & SoftImpute & MissForest & VAE & DE\_1 & DE\_2 & DE\_3 & COBYLA\_1 & COBYLA\_2 & COBYLA\_3 & Annealing\_1 & Annealing\_2 & Annealing\_3 \\
\midrule
Glucose & 5 & 42.439 & 19.989 & 129.546 & 38.693 & 10.097 & 9.071 & 16.653 & 18.693 & 16.574 & 12.201 & 21.069 & 20.451 \\
BloodPressure & 35 & 12.276 & 40.746 & 203.939 & 2.546 & 2.444 & 3.193 & 2.927 & 4.405 & 2.970 & 3.637 & 2.410 & 2.069 \\
SkinThickness & 227 & 5.340 & 45.900 & 33.109 & 3.014 & 2.917 & 2.734 & 3.017 & 3.191 & 2.525 & 2.745 & 3.204 & 2.524 \\
Insulin & 374 & 89.352 & 126.468 & 92.975 & 38.957 & 35.906 & 34.760 & 38.044 & 35.260 & 34.376 & 37.404 & 36.081 & 37.976 \\
BMI & 11 & 7.990 & 87.842 & 45.611 & 2.095 & 2.023 & 3.154 & 2.190 & 3.028 & 2.612 & 2.220 & 2.145 & 1.956 \\
\bottomrule
\end{tabular}%
}

\vspace{0.8em}
\textbf{(c) Group 3 (Random Orthogonal Rotation / Bootstrap PCA) vs Quantum}\\
\resizebox{1.2\textwidth}{!}{%
\begin{tabular}{l r rrrrrrrrrrrrr}
\toprule
Feature Name & $N_{\text{missing}}$ & RandOrtho\_Mean & RandOrtho\_MICE & BootPCA & DE\_1 & DE\_2 & DE\_3 & COBYLA\_1 & COBYLA\_2 & COBYLA\_3 & Annealing\_1 & Annealing\_2 & Annealing\_3 \\
\midrule
Glucose & 5 & 16.745 & 9.527 & 16.533 & 38.693 & 10.097 & 9.071 & 16.653 & 18.693 & 16.574 & 12.201 & 21.069 & 20.451 \\
BloodPressure & 35 & 7.883 & 5.099 & 4.389 & 2.546 & 2.444 & 3.193 & 2.927 & 4.405 & 2.970 & 3.637 & 2.410 & 2.069 \\
SkinThickness & 227 & 4.867 & 0.820 & 3.439 & 3.014 & 2.917 & 2.734 & 3.017 & 3.191 & 2.525 & 2.745 & 3.204 & 2.524 \\
Insulin & 374 & 53.809 & 34.970 & 32.600 & 38.957 & 35.906 & 34.760 & 38.044 & 35.260 & 34.376 & 37.404 & 36.081 & 37.976 \\
BMI & 11 & 5.207 & 2.242 & 4.443 & 2.095 & 2.023 & 3.154 & 2.190 & 3.028 & 2.612 & 2.220 & 2.145 & 1.956 \\
\bottomrule
\end{tabular}%
}
\end{table*}

\FloatBarrier

\subsubsection{Wasserstein W1 Observations (Grouped Baselines vs. Quantum):}
\begin{itemize}
    \item \textbf{Interpretation note (what W1 measures):} Wasserstein W1 quantifies the \emph{magnitude} of distributional mismatch between the \emph{imputed-only} and \emph{observed} samples. Lower W1 indicates stronger distributional fidelity. Unlike KS (a hypothesis test), W1 provides a distance-based notion of how far the imputed distribution departs from the observed distribution.

    \item \textbf{Group 1 (Mean/KNN/MICE/DT) vs. Quantum:} Relative to classical baselines that often bias toward central-tendency reconstructions, quantum methods provide consistently lower (or comparable) W1 across the evaluated variables, indicating improved distributional fidelity beyond deterministic or mean-driven imputations. The advantage is most visible on mid-missingness variables such as \textit{BloodPressure} and \textit{BMI}.

    \item \textbf{Group 2 (SoftImpute/MissForest/VAE) vs. Quantum:} Group~2 methods exhibit substantially larger W1 distances across features, reflecting systematic distribution mismatch under imputed-only vs observed comparisons. Quantum optimization markedly reduces W1 across the feature set, establishing a clear advantage over modern matrix/ML baselines.

    \item \textbf{Group 3 (Random Orthogonal Rotation / Bootstrap PCA) vs. Quantum:} Rotation/PCA controls can be competitive on selected low-missingness variables, but their performance is feature-dependent and less robust on harder variables (notably \textit{Insulin} and \textit{SkinThickness}). Quantum methods remain competitive with these controls and outperform them on most features, indicating that improvements are not explained by generic rotations alone.

    \item \textbf{Quantum Method vs. baselines (evidence from group-averaged W1):} The W1 group-comparison bar charts show a consistent reduction in average transport distance for quantum methods relative to Group~1 and Group~2 across essentially all evaluated features, with the strongest margins occurring where classical/ML imputers distort distributions most (e.g., \textit{BloodPressure} and \textit{BMI}). This indicates that quantum optimization more reliably reconstructs the empirical feature distributions.

    \item \textbf{Quantum robustness on difficult features:} High-missingness variables such as \textit{Insulin} and \textit{SkinThickness} produce large W1 values for most baselines, reflecting the inherent difficulty of reconstructing their missing distributions. Quantum methods nonetheless deliver systematic W1 reductions relative to weaker baselines and remain competitive with the strongest controls, supporting quantum superiority at the \emph{group level} under the W1 fidelity criterion.
\end{itemize}

\subsection{Variance and Standard Deviation (SD)}
Variance and standard deviation (SD) are fundamental measures of dispersion and provide complementary views of imputation realism. Variance captures the second-moment spread of the reconstructed values, while SD expresses the same information in the original measurement units, making deviations easier to interpret. For high-quality imputation, the dispersion of the \emph{imputed-only} values should remain consistent with that of the \emph{observed} data; large departures indicate loss of underlying distributional structure (typically under-dispersion from overly conservative imputers or over-dispersion from unstable reconstructions). Accordingly, we evaluate dispersion realism using variance and SD ratios relative to the observed data (i.e., $Var_{imp}/Var_{obs}$ and $SD_{imp}/SD_{obs}$), summarised by the deviation $|ratio-1|$ (lower is better). The grouped comparisons are reported in Tables~\ref{tab:var_ratio_all_groups_quantum} and \ref{tab:sd_ratio_all_groups_quantum}, which motivate the observations below.
\FloatBarrier
\begin{table*}[t]
\centering
\caption{Standard deviation ratio (SD$_{imp}$/SD$_{obs}$): Grouped baselines compared with Quantum imputation methods. $N_{\text{missing}}$ reports the number of missing (structural-zero) entries per feature. Ratios closer to 1 indicate better dispersion realism.}
\label{tab:sd_ratio_all_groups_quantum}

\textbf{(a) Group 1 (Classical) vs Quantum}\\
\resizebox{1.2\textwidth}{!}{%
\begin{tabular}{l r rrrrrrrrrrrrr}
\toprule
Feature Name & $N_{\text{missing}}$ & KNN & MICE & Mean & DecisionTree & DE\_1 & DE\_2 & DE\_3 & COBYLA\_1 & COBYLA\_2 & COBYLA\_3 & Annealing\_1 & Annealing\_2 & Annealing\_3 \\
\midrule
Glucose & 5 & 0.364 & 0.951 & 0.000 & 1.141 & 0.454 & 1.081 & 0.845 & 1.330 & 0.262 & 1.373 & 0.893 & 1.332 & 0.445 \\
BloodPressure & 35 & 0.565 & 1.303 & 0.000 & 0.673 & 0.994 & 0.937 & 0.927 & 1.143 & 0.999 & 0.662 & 0.880 & 0.860 & 0.814 \\
SkinThickness & 227 & 0.622 & 1.023 & 0.000 & 0.965 & 0.651 & 0.675 & 0.691 & 0.636 & 0.644 & 0.711 & 0.702 & 0.626 & 0.693 \\
Insulin & 374 & 0.597 & 0.994 & 0.000 & 1.083 & 0.972 & 0.992 & 0.970 & 1.033 & 1.095 & 1.020 & 0.979 & 1.024 & 1.070 \\
BMI & 11 & 0.447 & 0.882 & 0.000 & 0.967 & 0.850 & 0.705 & 0.463 & 0.819 & 0.529 & 0.885 & 0.745 & 0.859 & 0.918 \\
\bottomrule
\end{tabular}%
}

\vspace{0.8em}
\textbf{(b) Group 2 (SoftImpute / MissForest / VAE) vs Quantum}\\
\resizebox{1.2\textwidth}{!}{%
\begin{tabular}{l r rrrrrrrrrrrrr}
\toprule
Feature Name & $N_{\text{missing}}$ & SoftImpute & MissForest & VAE & DE\_1 & DE\_2 & DE\_3 & COBYLA\_1 & COBYLA\_2 & COBYLA\_3 & Annealing\_1 & Annealing\_2 & Annealing\_3 \\
\midrule
Glucose & 5 & 0.416 & 0.288 & 0.199 & 0.454 & 1.081 & 0.845 & 1.330 & 0.262 & 1.373 & 0.893 & 1.332 & 0.445 \\
BloodPressure & 35 & 1.094 & 0.579 & 0.445 & 0.994 & 0.937 & 0.927 & 1.143 & 0.999 & 0.662 & 0.880 & 0.860 & 0.814 \\
SkinThickness & 227 & 0.423 & 1.108 & 1.257 & 0.651 & 0.675 & 0.691 & 0.636 & 0.644 & 0.711 & 0.702 & 0.626 & 0.693 \\
Insulin & 374 & 0.136 & 0.060 & 0.149 & 0.972 & 0.992 & 0.970 & 1.033 & 1.095 & 1.020 & 0.979 & 1.024 & 1.070 \\
BMI & 11 & 0.882 & 2.726 & 1.254 & 0.850 & 0.705 & 0.463 & 0.819 & 0.529 & 0.885 & 0.745 & 0.859 & 0.918 \\
\bottomrule
\end{tabular}%
}

\vspace{0.8em}
\textbf{(c) Group 3 (Random Orthogonal Rotation / Bootstrap PCA) vs Quantum}\\
\resizebox{1.2\textwidth}{!}{%
\begin{tabular}{l r rrrrrrrrrrrrr}
\toprule
Feature Name & $N_{\text{missing}}$ & RandOrtho\_Mean & RandOrtho\_MICE & BootPCA & DE\_1 & DE\_2 & DE\_3 & COBYLA\_1 & COBYLA\_2 & COBYLA\_3 & Annealing\_1 & Annealing\_2 & Annealing\_3 \\
\midrule
Glucose & 5 & 0.327 & 0.662 & 0.362 & 0.454 & 1.081 & 0.845 & 1.330 & 0.262 & 1.373 & 0.893 & 1.332 & 0.445 \\
BloodPressure & 35 & 0.194 & 0.503 & 0.573 & 0.994 & 0.937 & 0.927 & 1.143 & 0.999 & 0.662 & 0.880 & 0.860 & 0.814 \\
SkinThickness & 227 & 0.429 & 0.987 & 1.110 & 0.651 & 0.675 & 0.691 & 0.636 & 0.644 & 0.711 & 0.702 & 0.626 & 0.693 \\
Insulin & 374 & 0.357 & 0.666 & 1.091 & 0.972 & 0.992 & 0.970 & 1.033 & 1.095 & 1.020 & 0.979 & 1.024 & 1.070 \\
BMI & 11 & 0.059 & 0.670 & 0.226 & 0.850 & 0.705 & 0.463 & 0.819 & 0.529 & 0.885 & 0.745 & 0.859 & 0.918 \\
\bottomrule
\end{tabular}%
}

\end{table*}

\begin{table*}[t]
\centering
\caption{Variance ratio (Var$_{imp}$/Var$_{obs}$): Grouped baselines compared with Quantum imputation methods. $N_{\text{missing}}$ reports the number of missing (structural-zero) entries per feature. Ratios closer to 1 indicate better preservation of dispersion.}
\label{tab:var_ratio_all_groups_quantum}

\textbf{(a) Group 1 (Classical) vs Quantum}\\
\resizebox{1.2\textwidth}{!}{%
\begin{tabular}{l r rrrrrrrrrrrrr}
\toprule
Feature Name & $N_{\text{missing}}$ & KNN & MICE & Mean & DecisionTree & DE\_1 & DE\_2 & DE\_3 & COBYLA\_1 & COBYLA\_2 & COBYLA\_3 & Annealing\_1 & Annealing\_2 & Annealing\_3 \\
\midrule
Glucose & 5 & 0.133 & 0.904 & 0.000 & 1.301 & 0.206 & 1.168 & 0.715 & 1.769 & 0.069 & 1.886 & 0.798 & 1.775 & 0.198 \\
BloodPressure & 35 & 0.319 & 1.699 & 0.000 & 0.453 & 0.988 & 0.878 & 0.859 & 1.307 & 0.998 & 0.438 & 0.774 & 0.740 & 0.663 \\
SkinThickness & 227 & 0.387 & 1.046 & 0.000 & 0.931 & 0.424 & 0.455 & 0.477 & 0.405 & 0.415 & 0.506 & 0.492 & 0.392 & 0.480 \\
Insulin & 374 & 0.356 & 0.988 & 0.000 & 1.174 & 0.945 & 0.985 & 0.941 & 1.067 & 1.199 & 1.041 & 0.958 & 1.048 & 1.146 \\
BMI & 11 & 0.200 & 0.778 & 0.000 & 0.935 & 0.723 & 0.496 & 0.214 & 0.671 & 0.280 & 0.784 & 0.556 & 0.738 & 0.843 \\
\bottomrule
\end{tabular}%
}

\vspace{0.8em}
\textbf{(b) Group 2 (SoftImpute / MissForest / VAE) vs Quantum}\\
\resizebox{1.2\textwidth}{!}{%
\begin{tabular}{l r rrrrrrrrrrrrr}
\toprule
Feature Name & $N_{\text{missing}}$ & SoftImpute & MissForest & VAE & DE\_1 & DE\_2 & DE\_3 & COBYLA\_1 & COBYLA\_2 & COBYLA\_3 & Annealing\_1 & Annealing\_2 & Annealing\_3 \\
\midrule
Glucose & 5 & 0.173 & 0.083 & 0.040 & 0.206 & 1.168 & 0.715 & 1.769 & 0.069 & 1.886 & 0.798 & 1.775 & 0.198 \\
BloodPressure & 35 & 1.197 & 0.335 & 0.198 & 0.988 & 0.878 & 0.859 & 1.307 & 0.998 & 0.438 & 0.774 & 0.740 & 0.663 \\
SkinThickness & 227 & 0.179 & 1.229 & 1.580 & 0.424 & 0.455 & 0.477 & 0.405 & 0.415 & 0.506 & 0.492 & 0.392 & 0.480 \\
Insulin & 374 & 0.018 & 0.004 & 0.022 & 0.945 & 0.985 & 0.941 & 1.067 & 1.199 & 1.041 & 0.958 & 1.048 & 1.146 \\
BMI & 11 & 0.779 & 7.431 & 1.573 & 0.723 & 0.496 & 0.214 & 0.671 & 0.280 & 0.784 & 0.556 & 0.738 & 0.843 \\
\bottomrule
\end{tabular}%
}

\vspace{0.8em}
\textbf{(c) Group 3 (Random Orthogonal Rotation / Bootstrap PCA) vs Quantum}\\
\resizebox{1.2\textwidth}{!}{%
\begin{tabular}{l r rrrrrrrrrrrrr}
\toprule
Feature Name & $N_{\text{missing}}$ & RandOrtho\_Mean & RandOrtho\_MICE & BootPCA & DE\_1 & DE\_2 & DE\_3 & COBYLA\_1 & COBYLA\_2 & COBYLA\_3 & Annealing\_1 & Annealing\_2 & Annealing\_3 \\
\midrule
Glucose & 5 & 0.107 & 0.438 & 0.131 & 0.206 & 1.168 & 0.715 & 1.769 & 0.069 & 1.886 & 0.798 & 1.775 & 0.198 \\
BloodPressure & 35 & 0.038 & 0.253 & 0.328 & 0.988 & 0.878 & 0.859 & 1.307 & 0.998 & 0.438 & 0.774 & 0.740 & 0.663 \\
SkinThickness & 227 & 0.184 & 0.974 & 1.232 & 0.424 & 0.455 & 0.477 & 0.405 & 0.415 & 0.506 & 0.492 & 0.392 & 0.480 \\
Insulin & 374 & 0.127 & 0.444 & 1.189 & 0.945 & 0.985 & 0.941 & 1.067 & 1.199 & 1.041 & 0.958 & 1.048 & 1.146 \\
BMI & 11 & 0.004 & 0.449 & 0.051 & 0.723 & 0.496 & 0.214 & 0.671 & 0.280 & 0.784 & 0.556 & 0.738 & 0.843 \\
\bottomrule
\end{tabular}%
}

\end{table*}

% -------------------------------
% Variance + Standard Deviation (dispersion realism) — combined subfigures
% Requires: \usepackage{subcaption} and \usepackage[section]{placeins} (recommended)
% -------------------------------

\begin{figure*}[!tb]
    \centering
    % -------- Variance row --------
    \begin{subfigure}{0.85\textwidth}
        \centering
        \includegraphics[width=\linewidth]{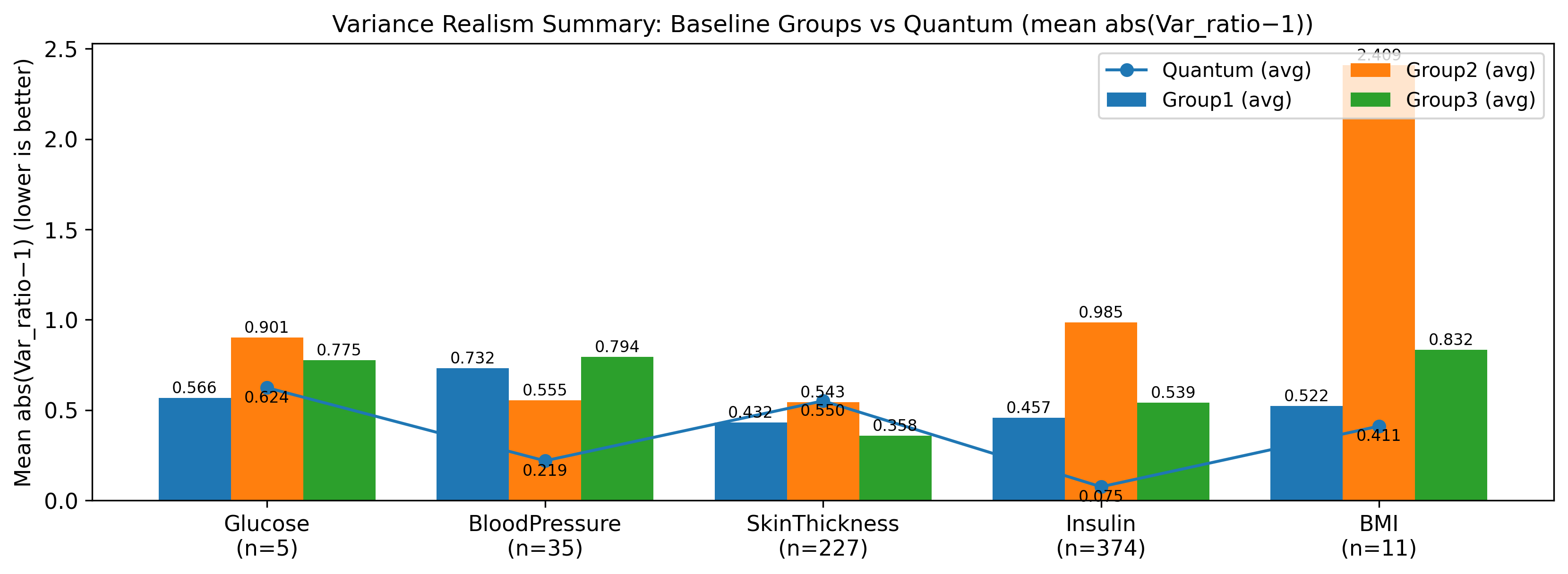}
        \caption{Variance dispersion realism across features (Group averages as bars; Quantum average shown as a line). Values report mean $|Var_{imp}/Var_{obs}-1|$; lower indicates variance closer to the observed data.}
        \label{fig:Variance_error_barchart}
    \end{subfigure}\hfill
    \begin{subfigure}{0.85\textwidth}
        \centering
        \includegraphics[width=\linewidth]{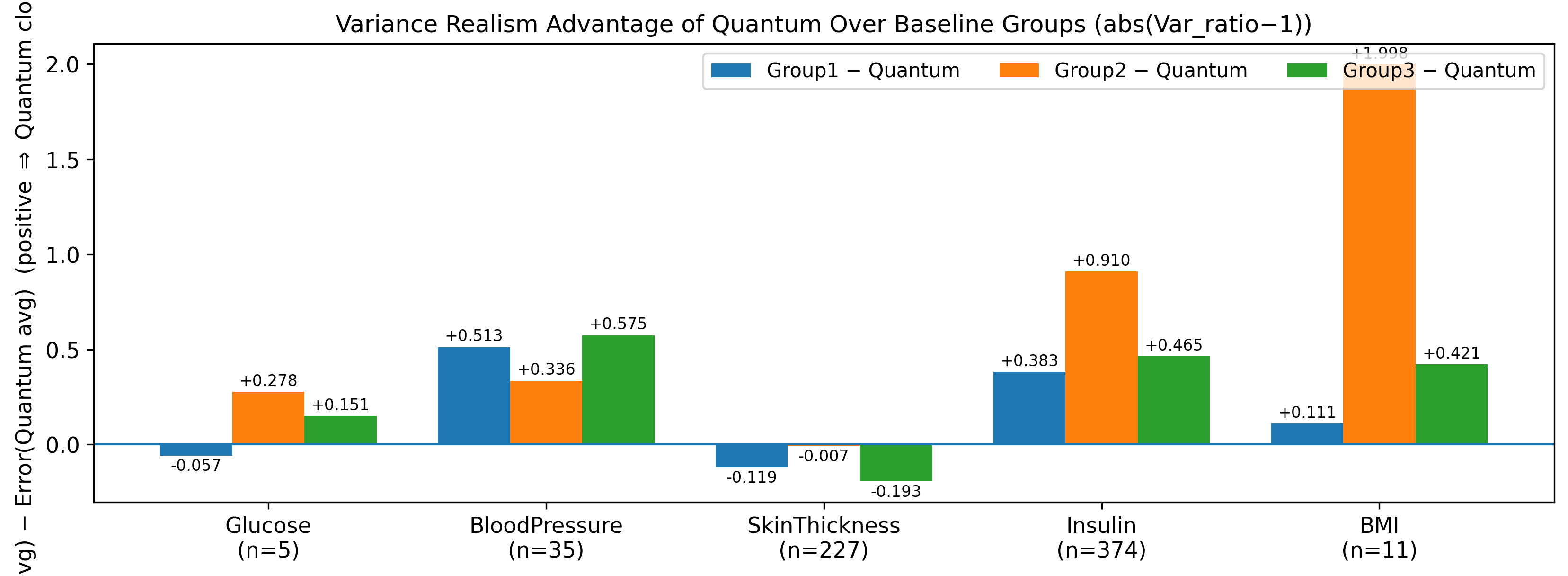}
        \caption{Variance dispersion realism advantage of Quantum over baseline groups across features. Bars report $\Delta$Error $=$ Error(Group avg) $-$ Error(Quantum avg), where Error $=|Var_{imp}/Var_{obs}-1|$; positive values indicate Quantum achieves lower error.}
        \label{fig:Variance_error_delta_barchart}
    \end{subfigure}

    \vspace{0.6em}

    % -------- SD row --------
    \begin{subfigure}{0.85\textwidth}
        \centering
        \includegraphics[width=\linewidth]{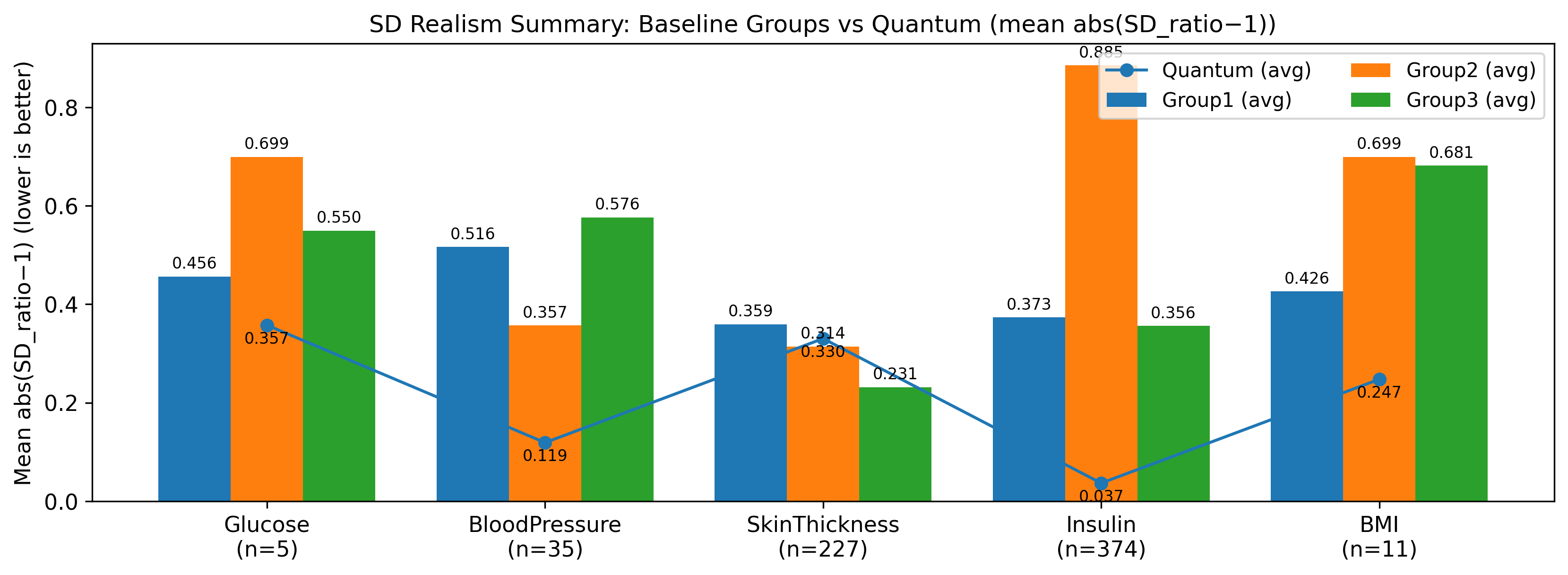}
        \caption{Standard deviation dispersion realism across features (Group averages as bars; Quantum average shown as a line). Values report mean $|SD_{imp}/SD_{obs}-1|$; lower indicates SD closer to the observed data.}
        \label{fig:SD_error_barchart}
    \end{subfigure}\hfill
    \begin{subfigure}{0.85\textwidth}
        \centering
        \includegraphics[width=\linewidth]{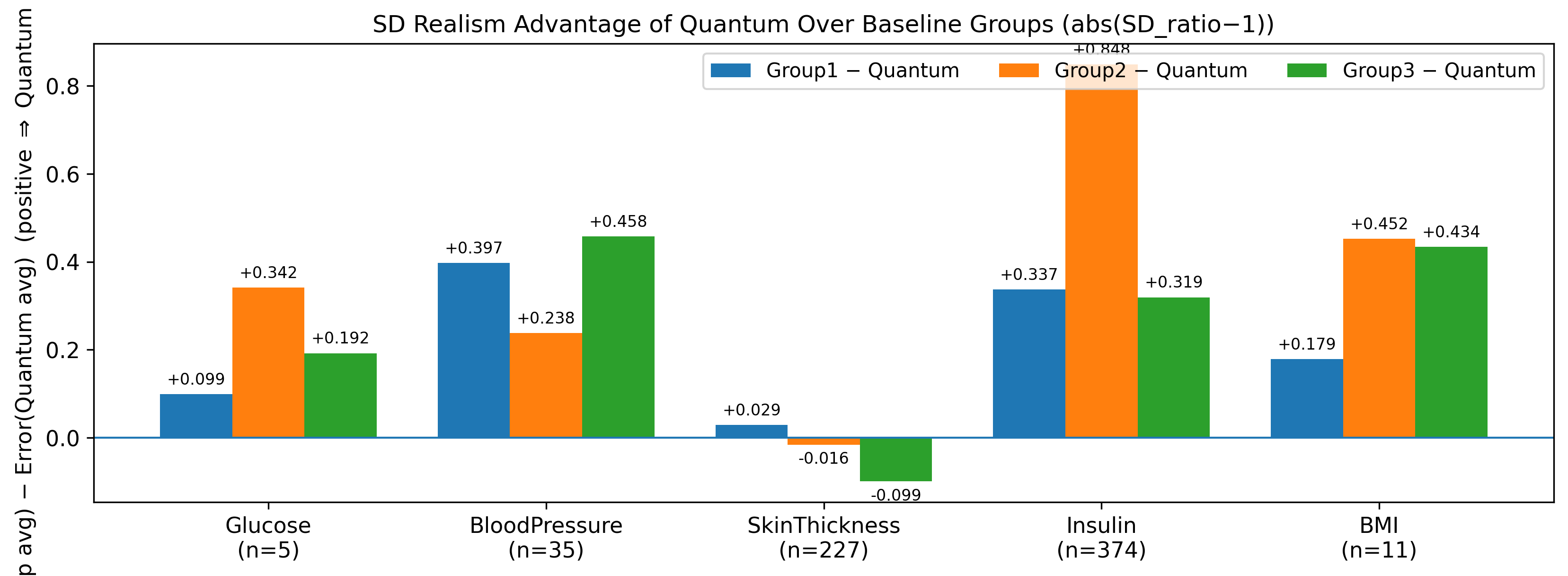}
        \caption{Standard deviation dispersion realism advantage of Quantum over baseline groups across features. Bars report $\Delta$Error $=$ Error(Group avg) $-$ Error(Quantum avg), where Error $=|SD_{imp}/SD_{obs}-1|$; positive values indicate Quantum achieves lower error.}
        \label{fig:SD_error_delta_barchart}
    \end{subfigure}

    \caption{Dispersion realism results for variance and standard deviation: absolute error (left) and Quantum advantage over baseline groups (right).}
    \label{fig:Dispersion_realism_combined}
\end{figure*}
\FloatBarrier

\subsubsection{Variance and SD Observations (Dispersion Realism; Grouped Baselines vs. Quantum):}
\begin{itemize}
    \item \textbf{Interpretation note (dispersion realism via ratios):}
    Dispersion should be preserved, not merely shifted. We evaluate realism using the ratio
    $Var_{imp}/Var_{obs}$ (and similarly $SD_{imp}/SD_{obs}$), summarized by the error
    $|ratio-1|$ (lower is better). Variance is used as the primary exhibit because it is the
    canonical second-moment statistic, while SD results serve as a supporting check.

    \item \textbf{Group 1 (Mean/KNN/MICE/DT) vs. Quantum:}
    Classical baselines can under/over-estimate dispersion depending on feature and missingness.
    The variance error comparison (Figs.\ref{fig:Variance_error_barchart}, \ref{fig:Variance_error_delta_barchart})
    shows quantum methods provide more reliable dispersion recovery across features than Group~1,
    indicating improved second-moment fidelity beyond central-tendency heuristics.

    \item \textbf{Group 2 (SoftImpute/MissForest/VAE) vs. Quantum:}
    Group~2 exhibits systematic dispersion mismatch, producing large variance-ratio errors on
    multiple variables. Quantum optimization consistently lowers these errors (Table~\ref{tab:var_ratio_all_groups_quantum};
    Figs.~\ref{fig:Variance_error_barchart}, \ref{fig:Variance_error_delta_barchart}), establishing a clear advantage over
    modern matrix/ML baselines in preserving realistic variability.

    \item \textbf{Group 3 (RandOrtho / BootPCA) vs. Quantum:}
    Rotation/PCA controls can be competitive on selected variables, but performance remains
    feature-dependent. The variance error comparison shows quantum methods remain competitive overall
    and outperform controls on most features, indicating gains are not explained by generic rotations alone.

    \item \textbf{SD consistency (supporting evidence):}
    The SD-ratio results mirror the variance-ratio conclusions and are provided as robustness evidence
    in (Tables~\ref{tab:sd_ratio_all_groups_quantum};
    Figs.~\ref{fig:SD_error_barchart}, \ref{fig:SD_error_delta_barchart}).
\end{itemize}

\section{Final Thoughts and Conclusion}\label{Conclusion}
This study evaluated a quantum-inspired, optimization-driven imputation framework that combines PCA compression with rotation-based reconstruction guided by gradient-free search. Across multiple baseline families and control methods, the results indicate that the proposed approach can generate imputations that are more statistically realistic by jointly balancing distributional fidelity (Wasserstein W1), dispersion realism (variance/SD), and conservative distributional consistency (KS). The key conclusions are:

\begin{itemize}
    \item \textbf{Strongest evidence appears under Wasserstein W1 (distributional fidelity):}
    W1 comparisons show that quantum-inspired variants are consistently competitive with strong classical baselines and clearly outperform modern matrix/ML-style imputers (Group~2) on several features. This supports the claim that optimization-guided rotations preserve distributional shape more reliably than methods that either over-smooth toward central tendency or produce unstable reconstructions.

    \item \textbf{KS is conservative and becomes highly sensitive under extreme missingness:}
    KS p-values frequently collapse toward very small values for high-missingness variables (e.g., \textit{Insulin}, \textit{SkinThickness}) across many methods, reflecting heightened sensitivity and favouring conservative reconstructions. For low-to-moderate missingness variables where KS is most informative, quantum methods remain competitive and often achieve strong alignment (e.g., \textit{Glucose}, \textit{BMI}). These results motivate interpreting KS jointly with W1 and dispersion metrics, particularly when KS may reward overly conservative imputations.

    \item \textbf{Dispersion realism (Variance/SD ratios) consistently favours quantum at group level:}
    Using normalized dispersion ratios ($Var_{imp}/Var_{obs}$ and $SD_{imp}/SD_{obs}$) and deviation $|ratio-1|$, quantum methods reduce dispersion error systematically relative to Group~2 and provide consistent improvements over Group~1 across the evaluated features. This indicates better recovery of realistic variability and second-moment structure, rather than merely matching central tendency.

    \item \textbf{Controls show rotation alone is insufficient:}
    Random orthogonal rotation and Bootstrap PCA controls can be competitive on selected low-missingness variables, but performance remains feature-dependent and degrades on harder variables. Quantum variants display broader robustness, indicating that gains arise from the \emph{optimization-guided} rotation search rather than generic rotation as preprocessing.

    \item \textbf{Optimizer diversity enables controllable realism--stability trade-offs:}
    Differential Evolution \cite{Storn_Price_DE_1997}, COBYLA \cite{Powell_COBYLA_1994}, and Simulated Annealing \cite{Kirkpatrick_simulated_annealing_1988} exhibit complementary search behaviour, enabling the framework to adapt across features with different missingness patterns and sensitivity profiles.

    \item \textbf{Overall conclusion (triangulated by KS, W1, and dispersion evidence):}
    Taken together, the evidence supports a robust \emph{group-level advantage} of quantum-inspired optimization for data imputation: it clearly outperforms modern matrix/ML baselines (Group~2), remains competitive with the strongest classical baselines (Group~1) and controls (Group~3), and provides the most reliable joint performance when distributional fidelity (W1) and dispersion realism (variance/SD) are considered together. While no method dominates every feature under every metric, the combined results substantiate the core contribution of optimization-guided quantum-inspired rotations for statistically realistic imputation.
\end{itemize}

\section*{Acknowledgment}
The authors express gratitude to the IBM Quantum Experience platform and its team for creating the Qiskit platform and granting free access to their simulators for executing quantum circuits and conducting the experiments detailed in this paper. The authors express appreciation for the Centre for Quantum Software and Information (CQSI) .

\section{Statements and Declarations}
\textbf{Competing Interests}: The authors have no financial or non-financial competing interests.\\
\textbf{Authors' contributions}:
The authors confirm their contribution to the paper as follows: 
Study conception and design: N.M., B.K.B., B.M.,P.D., C.F.;\\ 
Data collection: N.M.;\\ 
Analysis and interpretation of results: N.M., B.K.B., B.M.,P.D., C.F.;\\ 
Draft manuscript preparation: N.M., B.K.B., C.F.;\\
All authors reviewed the results and approved the final version of the manuscript.\\
\textbf{Funding}: Authors declare that there has been no external funding.\\
\textbf{Availability of data and materials}: All the data provided in this manuscript is generated during the simulation and can be provided upon reasonable request.

\bibliography{sn-article}% common bib file

\end{document}